\pdfoutput=1

 \documentclass[preprint,12pt]{elsarticle}

\usepackage{amssymb}
 \usepackage{amsthm}

\usepackage{algorithm}
\usepackage{algorithmic}
\usepackage{hyperref}
\usepackage{graphicx}
\usepackage{caption}
\usepackage{xcolor}
\usepackage{mathtools}
\usepackage{setspace}
\usepackage{ifpdf}

\journal{Neural Network}

\begin{document}

\begin{frontmatter}



\title{A Novel Model for Arbitration between Planning and Habitual Control Systems}


\author[rvt]{Farzaneh S. Fard}
\ead{fard@cs.dal.ca}

\author[rvt]{Thomas P. Trappenberg\corref{mycorrespondingauthor}}
\ead{tt@cs.dal.ca}

\address [rvt] {Faculty of Computer Science, Dalhousie University, Halifax, NS, Canada}
\cortext[mycorrespondingauthor]{\\ Corresponding author}

\begin{abstract}
It is well established that humans decision making and instrumental control uses multiple systems, some which use habitual action selection and some which require deliberate planning. Deliberate planning systems use predictions of action-outcomes using an internal model of the agent's environment, while habitual action selection systems learn to automate by repeating previously rewarded actions. Habitual control is computationally efficient but may be inflexible in changing environments. Conversely, deliberate planning may be computationally expensive, but flexible in dynamic environments. This paper proposes a general architecture comprising both control paradigms by introducing an arbitrator that controls which subsystem is used at any time. This system is implemented for a target-reaching task with a simulated two-joint robotic arm that comprises a supervised internal model and deep reinforcement learning. Through permutation of target-reaching conditions, we demonstrate that the proposed is capable of rapidly learning kinematics of the system without \textit{a priori} knowledge, and is robust to (A) changing environmental reward and kinematics, and (B) occluded vision. The arbitrator model is compared to exclusive deliberate planning with the internal model and exclusive habitual control instances of the model. 
The results show how such a model can harness the benefits of both systems, using fast decisions in reliable circumstances while optimizing performance in changing environments. In addition, the proposed model learns very fast. Finally, the system which includes internal models is able to reach the target under the visual occlusion, while the pure habitual system is unable to operate sufficiently under such conditions.
\end{abstract}


\end{frontmatter}


\section{Introduction}
Much of the current reinforcement learning (RL) literature is in the domain of model-free control. Such a learning agent learns a value function from interacting with the environment, usually updating a proposed value function from a temporal difference between the previous expectation and a new experience \citep{mnih2013playing, mnih2015human}. The value function is like a big lookup-table that can quickly supply evaluations for possible actions and hence provide guidance for actions in a fast and somewhat automated way. Such a decision system can be characterized as habitual. While habitual action selection takes time to learn and requires that similar previous situations have been encountered sufficiently, the advantage is that decisions and correspondingly actions can be generated in a timely manner. 

In contrast, a system that has some internal models of the environment can be used to derive a value function on demand for a specific situation. A prime example is a Markov decision problem where the reward function and transition function of the agent are known so that the Bellman equations can be used to calculate the optimal value function for every state action pair without interacting with the environment. Of course, this system requires learning of the internal models, which requires previous interactions with the environment. The learning of internal models can be achieved through some form of supervised learning. Once the models have been learned, the model-based system is able to calculate a value function on the fly. This resembles some form of internal deliberation. The advantage of such a system is its flexibility to new situations. However, deliberations take time so that a habitual system is preferable when it comes to situations that benefit from a higher degree of automation.

In this paper, we introduce a learning system that we call the {\bf Arbitrated Predictive Actor-Critic (APAC)} that combines a habitual reinforcement learning system with a supervised learning system of internal models. Most importantly, we introduce an arbitration system that mediates between their usage. 
We specifically discuss a situation in which both systems alone can solve an exemplary task so that we can study the consequences of their direct interactions in relation to their exclusive use. We show that this system is responsive to changes in the environment and that it can learn the reward function very fast. Our results demonstrate how the learning paradigm tend to rely on habits after learning the reward function. Our results are in line with evidence of human behaviour mentioned above.

\section{Theoretical Premises}
There is a lot of behavioural and neurophysicological evidence for different types of control systems in the brain that are usually termed habitual or model-free and goal-directed or model-based \citep{balleine1998goal, glascher2010states, daw2011model}. 
In particular, one control system associated with the prefrontal cortex \citep{miller2001integrative} predicts action-outcomes using an internal model of the agent's environment and hence can be associated with a control system that uses deliberative planning. We will use in this paper the term deliberative planning instead of goal directed model-based control to minimize the possible confusion between the models of the environment from the models of the value function. Another control pathway in the brain is associated with the dorsolateral area \citep{houk199513} learns to repeat previously rewarded actions that resemble a habitual system. 

Some research showed that the two different control systems are used in different situations and can be simultaneously active \citep{lengyel2007hippocampal, poldrack2001interactive}. For example, in the brain, the cortical system represents a generalized mapping of input distributions while hippocampal learning is an instance-based system \citep{lengyel2007hippocampal}. Moreover, when the model of the environment is known and there is sufficient time to plan, the best strategy is deliberate planning \citep{daw2005uncertainty}, but when the decision should be taken very fast the habitual controller is used \citep{blundell2016model}. Other work shows that cooperation and competition between different control systems in the brain happens especially when outcomes of each control system disagree, that is, if a deliberated planning task does not align with a habitual skill \citep{daw2005uncertainty,daw2011model,daw2013multiple,lee2014neural}. 

Moreover, feedback to learning systems can differ in different situations and can be provided from different modalities such as vision or auditory input. In machine learning, it is common to distinguish different learning paradigms. One is supervised learning where a teacher gives feedback from the knowledge of a desired behaviour. The system can be trained by comparing the actual output of a leaner to the desired output provided by the teacher. The teacher is basically a critic who can quantify an objective function that a leaner needs to optimize. Another slightly more general learning paradigm is reinforcement learning where the environment only provides some indication of value, often only after a series of actions have been chosen by a learning agent. Reinforcement learning is thus more general in that and it can be applied to a lot of more typical learning situations of an agent in an environment, and its basic form is about learning a critic. The subsystems in our model also align in our implementation with a supervised paradigm to learn internal models and a reinforcement learning paradigm to learn habitual control.   

\begin{figure*}
\centering    
\includegraphics[height=60mm]{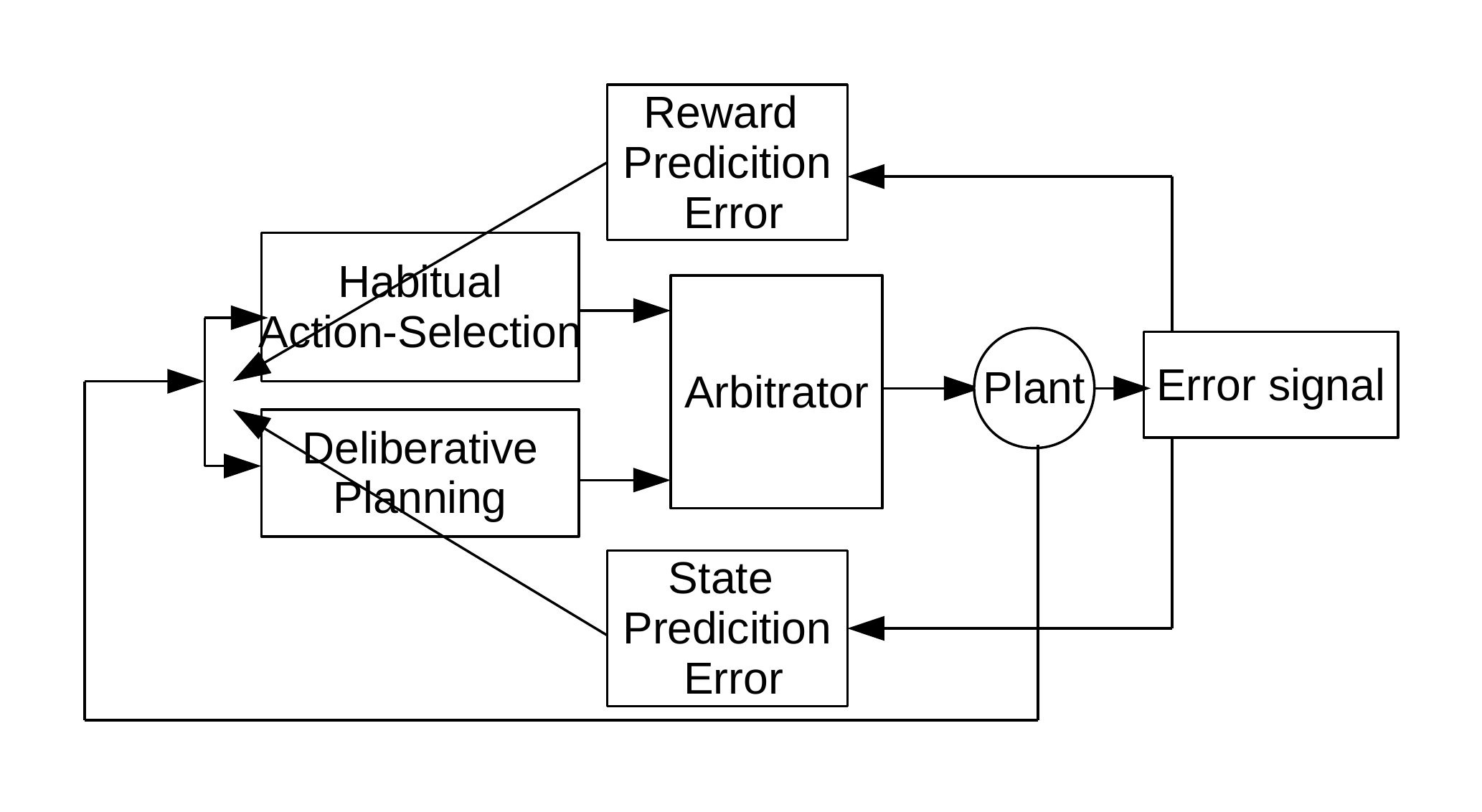}
\caption{\label{fig:general-view}General view of APAC model.}
\end{figure*}

Habitual reinforcement learning which is based on TD learning \citep{sutton1984temporal} has been very successful in explaining experimental evidence from the animal learning literature and dopamine-based learning in the brain \citep{barto199511,schultz1997neural}. Such models which have originally been formulated with tubular methods based on discrete state action spaces are now commonly combined with neural networks as a function approximator that broadens the range of practical applications to be solved using RL, especially for control problems with continuous states/actions spaces \citep{barto1990sequential, waltz1965heuristic}. Barto, Sutton and Anderson introduced the Actor-Critic architecture that was implemented by neural networks \citep{barto1983neuronlike}.  
Later, Barto \citep{barto199511} represented an adaptive critic which has similar behaviour to the dopamine neurons projection to the Striatum and frontal cortex. The adaptive critic uses the internal sensory information to learn an effective reinforcement signal.  

It has also long been hypothesized that the brain builds an internal representation of the world and its body \citep{miall1993cerebellum, miall1996forward, wolpert1998internal, kawato2003internal}, and evidence shows that "forward" and "inverse models" exist in the brain \citep{kawato2003internal, miall1993cerebellum}. The internal model is used to perform in the environment and learn a new task. Flanagan et al. \citep{flanagan1997role} showed that the internal model can predict the load force and the kinematics of a hand movement that depends on the load. Moreover, when learning how to use a new tool, humans make a transient change in the internal model of the arm as well as making an internal model of the tool \citep{kluzik2008reach}. Furthermore, imitation experiments show that a direct mapping happens between observation and the internal model \citep{iacoboni1999cortical}. Another advantage of having an internal representation is obtaining a reliable source of information for the agent to perform accurately even if there are no other sources of information (e.g. visual information) available \citep{wolpert1998internal, kawato2003internal}. 

In this paper, we propose a model to study the cooperation and competition between a habitual and planning-based control components with an arbitrator component. The general architecture of the proposed \textit{Arbitrated Predictive Actor-Critic} (APAC) is shown in Figure \ref{fig:general-view}. In this model, each control paradigm implies a specific type of teaching feedback. The deliberative planning controller incorporates internal models that are usually trained with supervised errors so that we consider here an explicit state predictions error. In contrast, the habitual action selection system is a common deep reinforcement learner which learns from reward prediction errors. The new component here is an arbitrator that mediates between these systems that can select the command given to the controlled system, the agent or in the plant in the common language of control theory. Of course, it is possible that both decision systems are trained with a combination of supervised and reinforcement learning, but this is not the crucial point in this paper. The model is designed to study how a combined control system behaves in different environmental situations. Indeed, in the following section we apply this general model to a specific motor control task in which both systems can be trained on the same feedback signal, but in which the execution would follow a habitual or planning implementation. 

\section{APAC for target reaching}
\begin{figure*}[t]
	\begin{center} \includegraphics[width=0.4\textwidth]{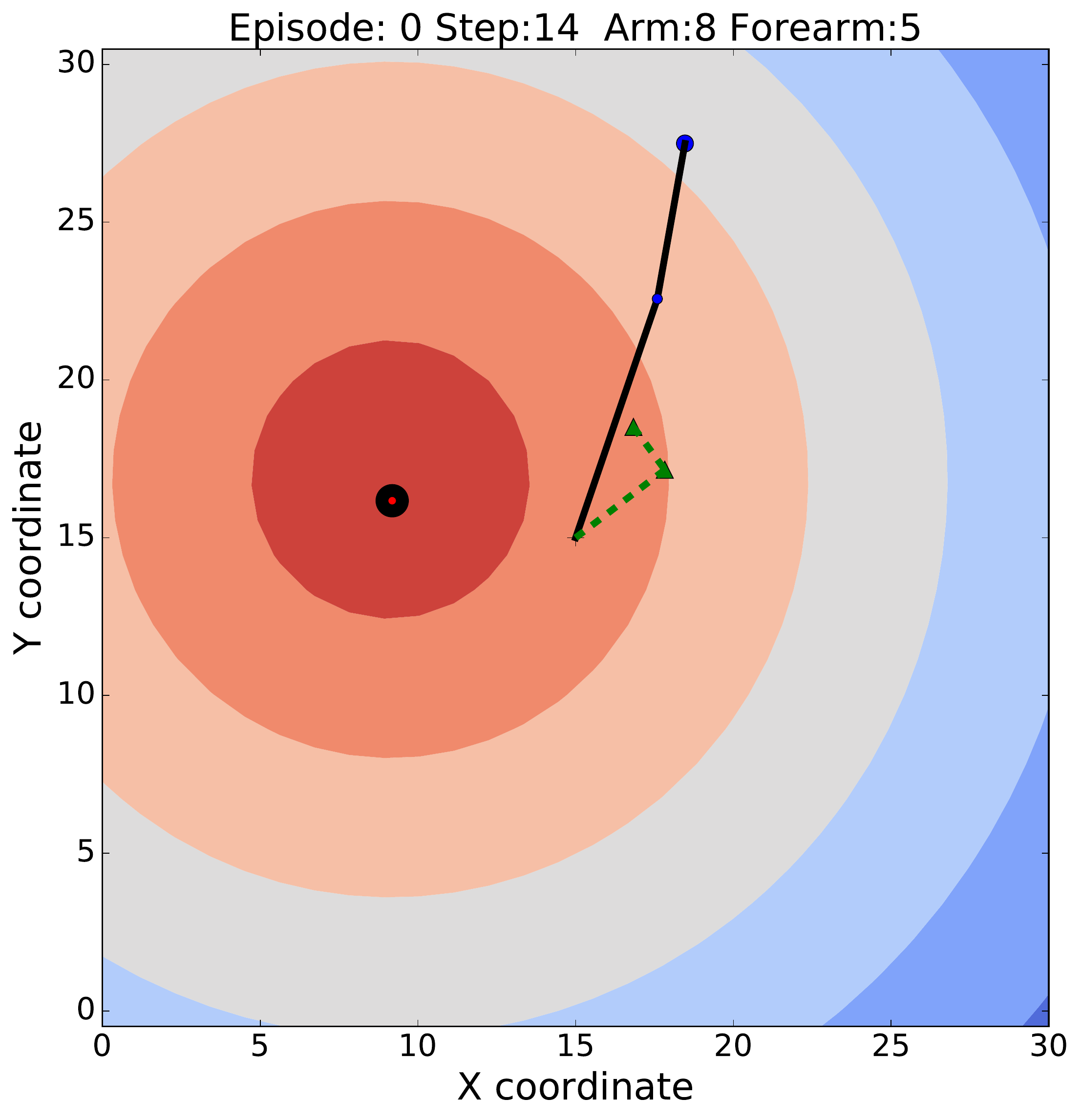} \end{center}
	\caption[Simulation environment]{\label{fig:RobotEnvironment1} The robotic arm set up in the environment. The solid black line indicates the robot arm. Blue and red circles display the end-effector and the target locations respectively. Colored contours illustrates reward function. Black circle shows the target zone. The dashed green line represents the internal model of the arm at very early stages of learning. }
\end{figure*}

In this section, we apply the APAC model to the motor control task of target reaching. We choose this task as it is a good example of a minimal control task while being complex enough to highlight the advantages and disadvantages of the two principle control architectures discussed in this paper. Target reaching lives in a continuous state and action space with 6 degrees of freedom when considering a shoulder and elbow yaw, pitch and roll, although we simplify this here even more to a 2-dimensional system with only one angle for the elbow and one the shoulder. Learning the reaching task in this 2D environment is learning a non-linear mapping function that maps joint angles of the robot arm onto a location of the end-effector in the environment.
An example image of our simulated robot arm is shown in Figure \ref{fig:RobotEnvironment1} with the black line. The contour plot shows the distance to the target while the dotted green line shows an internal model of the robot arm early in learning. 

\begin{figure*}
	\centering
	\includegraphics[width=125mm]{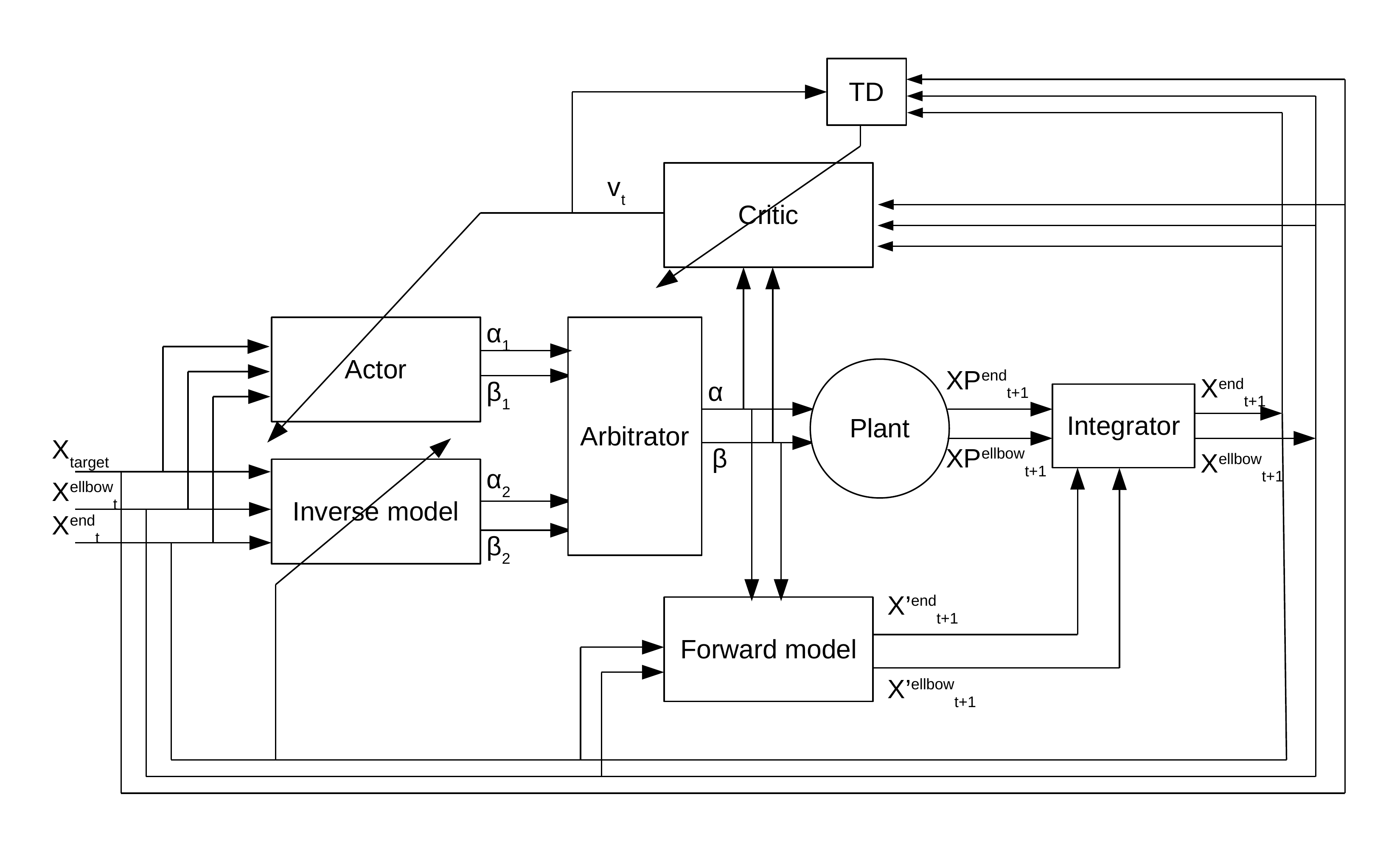}
	\caption[APAC]{Arbitrated Predictive Actor Critic: The actor receives the current state defined as the current location of the robot arm and the target location, and provides an action that is the change in angles for shoulder and elbow. The inverse model takes the current state and predicts another action. The output of the inverse model and the actor goes to the arbitrator. Then the arbitrator selects one of these actions. The output of the arbitrator along with the current state is the input to the critic from which the critic predicts a reward. The forward model receives the selected action and the current position of the agent and predicts the future state of the agent. The agent takes the selected actions and transfers to the next state. The predicted future state from the forward model is integrated with the estimation of the actual state of the plant after taking the action and obtains a new current state for the system. 
		\label{fig:APAC}}
\end{figure*}

The refined control architecture of our APAC model for the reaching tasks is shown in Figure \ref{fig:APAC}. For this application, the state is now defined as the position of the elbow, the end-effector, and the target. The planning component is now made out of a combination of deep forward and inverse models, while the habitual system is implemented as a deep actor-critic model. An integrator is used to derive the training signals that is used for the feedback. In the following, we specify each subsystem in detail.

\subsection{Habit learning control system}
The habitual controller is implemented as a deep deterministic policy gradient model (DDPG) following the work of Lillicrap et al.\citep{lillicrap2015continuous}. 
The arm position is given by the vector $X$ with two vector components, the position of the end-effector $X^{end}$ and the position of the elbow $X^{elbow}$. The arm position together with the target location $X_{target}$ defines the current state $s_t = [X^{end}, X^{elbow},X_{target}]$ of the agent. 

The critic $\mathcal{Q}(s_t,a_t;\theta^\mathcal{Q})$ is implemented as a deep neural network, where $s_t$ is the current state at time $t$, $a_t$ is the action taken at time $t$, and $\theta^\mathcal{Q}$ are the parameters of the critic network. The goal of the critic is to approximate the accumulation of the environmental reward (sometimes called return) that can be expected from a certain state action combination. The critic is learned through temporal difference (TD) learning \citep{sutton1988learning,schultz1997neural}.   
\begin{equation}
 \mathcal{Q}(s_t,a_t;\theta^\mathcal{Q})  \leftarrow \mathcal{Q}(s_t,a_t;\theta^\mathcal{Q})+l_1 \delta,
 \label{eq:TD}
\end{equation}
\begin{equation}
\delta = \underbrace{r_t + \gamma \max_{a'} \mathcal{Q'}(s_{t+1},a';\theta^{\mathcal{Q}} )}_\textrm{estimated reward}- \underbrace{\mathcal{Q}(s_t,a_t;\theta^\mathcal{Q})}_\textrm{actual reward},
\label{eq:rpe}
\end{equation}
where $l_1$ is the learning rate of the critic network, $r_t$ is the actual immediate reward received from the environment at time $t$, $\gamma$ is a discount factor, and $\mathcal{Q'}$ represents the estimation of the value of a state-action pair. As in DDPG, we use the main network for training but we use a \textit{target network}, which is a less frequently updated copy of the main network to avoid oscillation, for predicting. DDPG actually has two target networks, one for the critic network and one for the actor network. We follow directly the smooth update for the target networks as in DDPG \citealp{lillicrap2015continuous},
\begin{equation}
\theta' = \theta'\times (1-\tau) + \theta \times \tau,
\label{smoothUpdate}
 \end{equation}
with change parameter $\tau \ll 1 $. The parameters $\theta'$ represents the target network parameters, and $\theta$ is the parameter of the main network, either the actor or the critic.  

The computation in Equation \ref{eq:rpe} is done in the TD component. To train the critic using the TD rule, the error needs to be back propagated through the critic. The error between estimated value and the actual value is used to compute the loss function of the critic (Eq. \ref{loss-critic}),
\begin{equation}
 L^\mathcal{Q} =1/N \sum(\delta)^2 .
 \label{loss-critic}
\end{equation}
DDPG takes advantage of the \textit{experience memory replay} which is a memory to restore and reuse past experiences. The memory replay is in form of $R(s_t,a_t,r_t,s_{t+1},T_t)$, where $s_t$ is the current state at time $t$, $a_t$ is the action taken at time $t$, $s_{t+1}$ is the next state, $r_t$ is the reward received at time $t$, and $T_t$ indicates whether the state at time $t+1$ is a terminal or not. The replay memory is a queue-like buffer with a finite size. The agent will forget older experiences and it will update its parameters based on its recent experiences. At each time step, a random batch of $N$ samples is selected from the experience memory replay, and this batch is used to train both the actor and the critic.

The actor ($\pi$) receives the current state ($s_t$) and predicts future actions to be taken ($a_t$).
\begin{equation}
\pi(s_t;\theta^\pi) = a_t,
\end{equation}
where $a_t = [\alpha_1, \beta_1]$, where $\alpha_1$ and $\beta_1$ are motor commands sent to the shoulder and elbow, respectively. The actor is implemented as a deep network where $\theta^\pi$ indicates the parameter of the actor network and is trained using the deterministic policy gradient method \citep{silver2014deterministic}. Note that the main actor network is used for training, however, the target network of the actor is used for the action prediction. 

The changes of the weights of the actor corresponded to the changes in expected reward with respect to the actor's parameters,
\begin{equation}
 \theta_t^\pi \leftarrow \theta_t^\pi +l_2 \frac{\partial \mathcal{Q}(s_t,a_t;\theta^\mathcal{Q})}{\partial\pi(s_t;\theta^\pi_t)}\frac{\partial\pi(s_t;\theta^\pi_t)}{\partial\theta_t^\pi},
 \label{eq:actor-update-rule20}
\end{equation} 
where $l_2$ here is the learning rate of the actor. 
The plant, which is the simulated arm in our example, takes the action and transitions to its new position $[Xp^{end}_{t+1},Xp^{elbow}_{t+1}]$, which forms the new state $s_{t+1}$ when combined with the target location. Like DDPG we apply noise to the environment using an Ornstein-Uhlenbeck process \citep{uhlenbeck1930theory} that results in new samples. 
 
\subsection{Internal models for planning}
For the planning controller, we need to learn the transition function of the plant to build the model of the environment. Here we use supervised learning to learn the internal representation of the agent. More specifically, we used a supervised learning controller that uses past experiences to generalize an inverse model of the arm and a forward model of the arm. The training examples used in our implementation are obtained from the same experience replay memory that is used for the habitual controller.

A combination of a forward and an inverse model is used for planning the next actions. The forward model $f_F$ is a neural network that receives the current position of the arm $[X^{end}_t,X^{elbow}_t]$ and the action $a_t$ and predicts the future position of the arm $[X'^{end}_{t+1},X'^{elbow}_{t+1}]$. We can train the network from the discrepancy between the predicted future position $[XP^{end}_{t+1},XP^{ellbow}_{t+1}]$ and the actual position from visual feedback. For training we use the loss function
\begin{equation}
L^{f_F} = \frac{1}{N} \sum_t ([X'^{end}_{t+1},X'^{ellbow}_{t+1}] - [XP^{end}_{t+1},XP^{ellbow}_{t+1}])^2,
\label{loss-forward}
 \end{equation}
where $N$ is the number of selected samples in a batch of experiences stored in the replay memory. The size of the batch to train the forward model and the inverse model is the same as the one used for the actor and the critic.

An inverse model is another deep network, $f_I(s_t;\theta^{f_I})$. The aim of the inverse model is to provide a proper action to reach the target by minimizing the error between predicted action ($a'_t$) with the actual action taken ($a_t$) that transfers the agent from the current position to its next position. This network is then trained on the loss function: 
\begin{equation}
L^{f_I} = \frac{1}{N} \sum_t (a_t - a'_t)^2.
\label{loss-actor}
 \end{equation}

The aim of having the forward model is learning to predict future positions of the agent by taking specific actions. Such a model enables the agent to perform the task even with occluded vision.
When the inverse model has been trained well, it can be used to produce a suitable action to transfer the agent from its current state to the target location by replacing $X_{target}$ with $XP^{end}_t$. Hence, the inverse model can be trained with the input  $[X^{end}_{t-1},X^{elbow}_{t-1},XP^{end}_t]$ and predicting the proper actions on $[X^{end}_{t-1},X^{elbow}_{t-1},X_{target}]$. Note that $[X^{end}_{t-1},X^{elbow}_{t-1}]$ are part of states $s_t$ in the replay memory while $XP^{end}_t$ is taken from $s_{t+1}$ in the replay memory.

Another component of the overall system is "the integrator" module. In general, the integrator could be a Bayes filter such as a Kalman filter which estimates the best estimated position from the available information that combines an internal model prediction with external sensory feedback. Since we use a reliable visual feedback in our case study we simplify this to an integrator that passes the actual location of the plant in case visual information is available. With occluded vision, the prediction of the forward model is used as the estimated actual position of the agent. In our  previous work \citep{fard2015modeling}, we showed how to implement a Kalman Filter with Dynamic Neural Fields \citep{amari1977dynamics}. The integrator is the explicit critic in this example, which provides the state prediction error for the forward model (see Figure \ref{fig:general-view}). 

A training session of the system includes an infant phase that uses "motor babbling" \citep{iverson2004infant, von2004action, demiris2005motor, iverson2007relationship, caligiore2008using}. During the babbling phase, the plant produces random movements with random actions to produce actual samples to be stored in the experience memory. In the babbling phase, the actual position of the arm after taking an action is considered the target location. Therefore, all samples in this case reach the terminal state and will gain the maximum reward value. The babbling phase is used to provide valid examples to train both control systems.

\subsection{Arbitration between habitual and planning controllers}
A novel component of APAC is an arbitrator. The arbitrator receives action predictions from the deliberative planning module (the inverse model), and the habitual action selection module (the actor), and makes the final decision of which action to use. This selected action is transferred to the actuators of the plant to bring the agent into its new position resulting in a new state when combined with the target location. The arbitrator's decision is also fed into the forward model and the critic for training purposes. As in DDPG, noise from an Ornstein-Uhlenbeck process is added to both proposal actions provided by the inverse model and the actor. 

In our implementation of the APAC, we consider discrete action steps so that both controllers (habitual and planning) create actions at each step. However, it is known that the habitual controller is faster than deliberative planning. Therefore to imply the time constraint we set the arbitrator to give priority to the habitual controller. Moreover, the arbitrator is set to always take the action that is provided by the habitual system for the first two steps of every episode. However, from the third step on, the actor's prediction is taken if the habitual controller is reliable, meaning that the reward prediction error for the last experience is smaller than a threshold. We use $\delta < 1$ in the following experiments. Otherwise, the action from the inverse model is selected. It is, of course, possible to implement a more dynamic realization of such an arbitrator. For example, the threshold could itself be modulated according to the tasks and in this way produces a more rich speed-accuracy trade-off \citep{satel2005motivational}. However, the simple implementation discussed in this paper captures the minimal assumptions as outlined above and is sufficient for the following simulations.

\subsection{Experimental Conditions and Environment}
To tested the APAC model on a simulated robot arm with a target reaching task (Figure \ref{fig:RobotEnvironment1}), we simulated a two-joint robotic arm whose range of motion at each joint was constrained to 180 degrees. The arm’s motion was limited to a 2D plane of width 30 and height 30, upon which the arm’s "shoulder" was fixed in the centre (15,15). The initial length segment from the shoulder to the elbow was set to $l_1=5$, and the initial length of the lower segment ("hand" to "elbow") is $l_2=8$.

All experiments described herein had an episodic trial structure. At the beginning, the arm’s position was set to zero-degree angle at the shoulder and 180-degree angle at the elbow. Time was discretized in the simulations, and the learning agent was given only 30 action-steps per trial to achieve the designated goal. We define a "target zone" as a circle centered at the target location with a radius $r_{\rm target}=0.5$. The target is defined as "reached" once the robot arm is inside the target zone. If the goal was not reached within 30 time-steps, the trial was aborted and a new trial was started. 

Importantly, the reward function is defined as the negative Euclidean distance between the end-effector and the centre of the target area. This is for this example tasks the same information as is given to the supervised learner. This was deliberatively done so that the different systems are compared on the same feedback situation. It is possible to learn this task from much simpler feedback such as some reward if the target area is reached versus no reward otherwise, although this would then also need more time to train the habitual system.  This point of our study here is rather the direct comparison of decision components based on a value lookup versus learning internal models.      


Within this environment we define several conditions that defined the variety of the different target-reaching tasks studied here. These conditions include the target position ({\bf static target} vs. {\bf changing target} at each episode), kinematics (arm dimensions as {\bf static kinematics} vs. {\bf changing kinematics}), and vision ({\bf occluded vision} vs. {\bf perfect vision}). We tested all combinations of these factors.

Each experiment consisted of 1000 episodes of maximal 30 action steps each. 
In the static target condition, the target is initialized randomly and stays unchanged for all 1000 episodes. For the changing target condition, the target is located at a random location at the beginning of every episode. As discussed above, each episode was terminated when either (A) the target was reached, or (B) 30 time steps had elapsed. Targets were only placed within a reachable distance for the arm. The arm dimensions were kept fixed in the case of static kinematics; however, the length of both the upper and lower arm segments were increased by 0.001 at each time step for the changing kinematics condition. These changes were only started after the 100th episode of target reaching to provide some time for basic training. As already mentioned, an environmental noise was included in all experiments. In the occluded vision condition, the location of the arm and the target was unavailable for the agent during the movement. This task is also known as memory guided target reaching \citep{westwood2003no,heath2004control}.
We repeated all static/changing kinematics and static/changing target conditions with our proposed models for the reaching-target task in the occluded vision condition. 
To examined the generalization of the models under each condition, we trained each model when targets are located only in a specific area that represents 2/3 of whole reaching area, and tested with targets located in the other part of the environment, which is the rest 1/3 of the reaching area.

\section{Results}

We considered three versions of APAC that represent { a) exclusive habits}, { b) exclusive deliberate planning}, and { c) arbitration between habit and planning}. Exclusive habit is when the arbitrator is set to always pick the action from the habitual system. In this case, the APAC behaves exactly like DDPG. If the arbitrator always selects the action from the inverse model for each step, then the APAC becomes an exclusive deliberate planning controller which we call supervised predictive actor-critic (SPAC) \cite{fard2017anactorcritic}. The third model is when the APAC is able to arbitrate between the actions provided by the inverse model and the actor. 

For each condition, we trained 50 independent instances of each model for a total of 1000 episodes. At the end of the 1000th episode, all network parameters were frozen and no more training was applied. Subsequently, each of the independent model instances performed 100 target reaching episodes under the respective training conditions. In the case of the occluded vision, sensory input (i.e. visual target position) was initially presented at time step 0, and subsequently rendered unavailable. Since DDPG has no internal model and requires visual input throughout the task, we only compare APAC with SPAC in the experiments with occluded vision. All experiments were implemented and tested in Python (3.5) using the TensorFlow (1.3) package \cite{abadi2016tensorflow} on an NVIDIA GeForce GTX 960 graphical processing unit.



\begin{figure*}[!h]
\centering
\includegraphics[width=.24\textwidth]{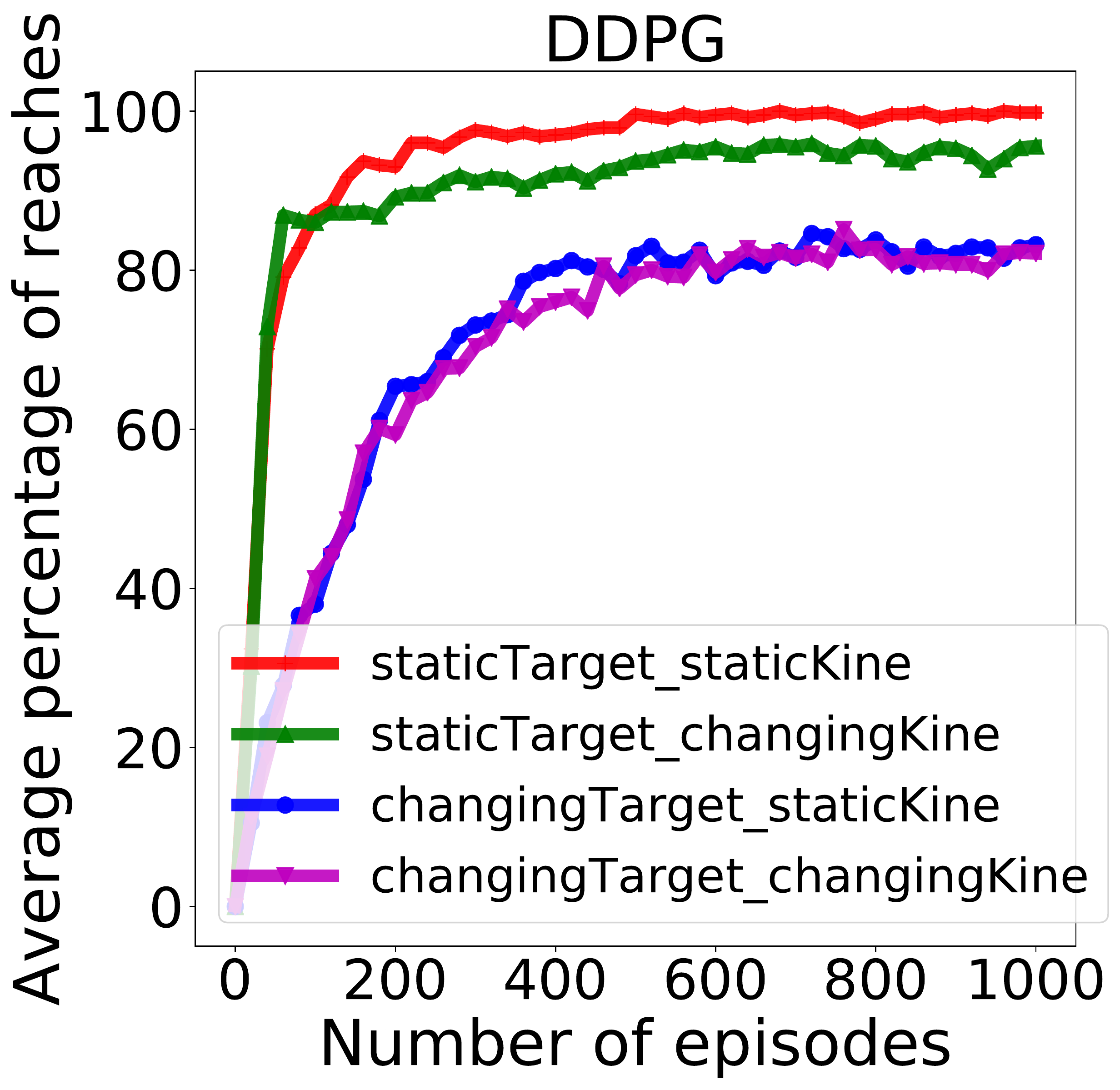}
\includegraphics[width=.24\textwidth]{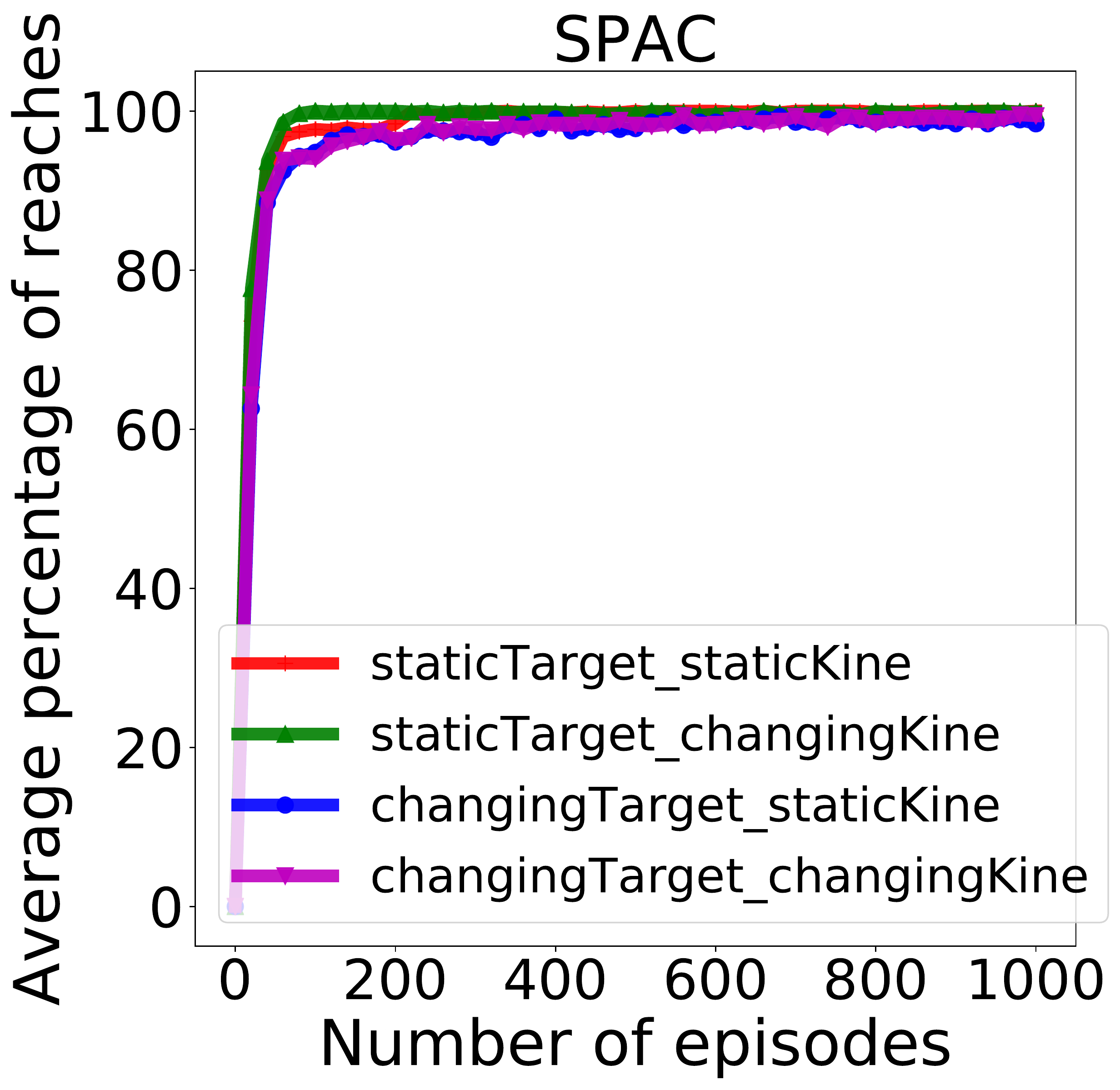}
\includegraphics[width=.24\textwidth]{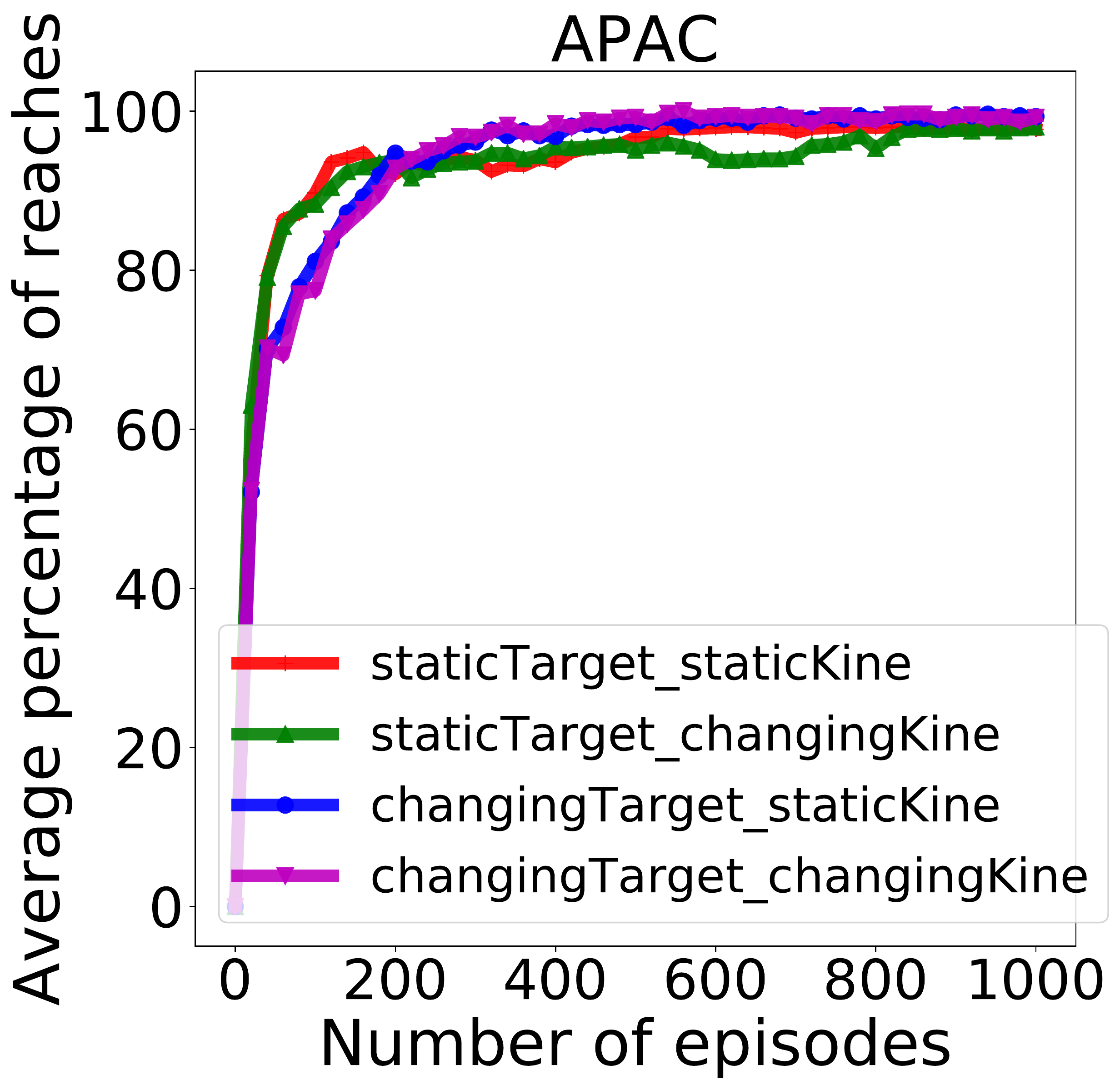}
\\ 
\includegraphics[width=.24\textwidth]{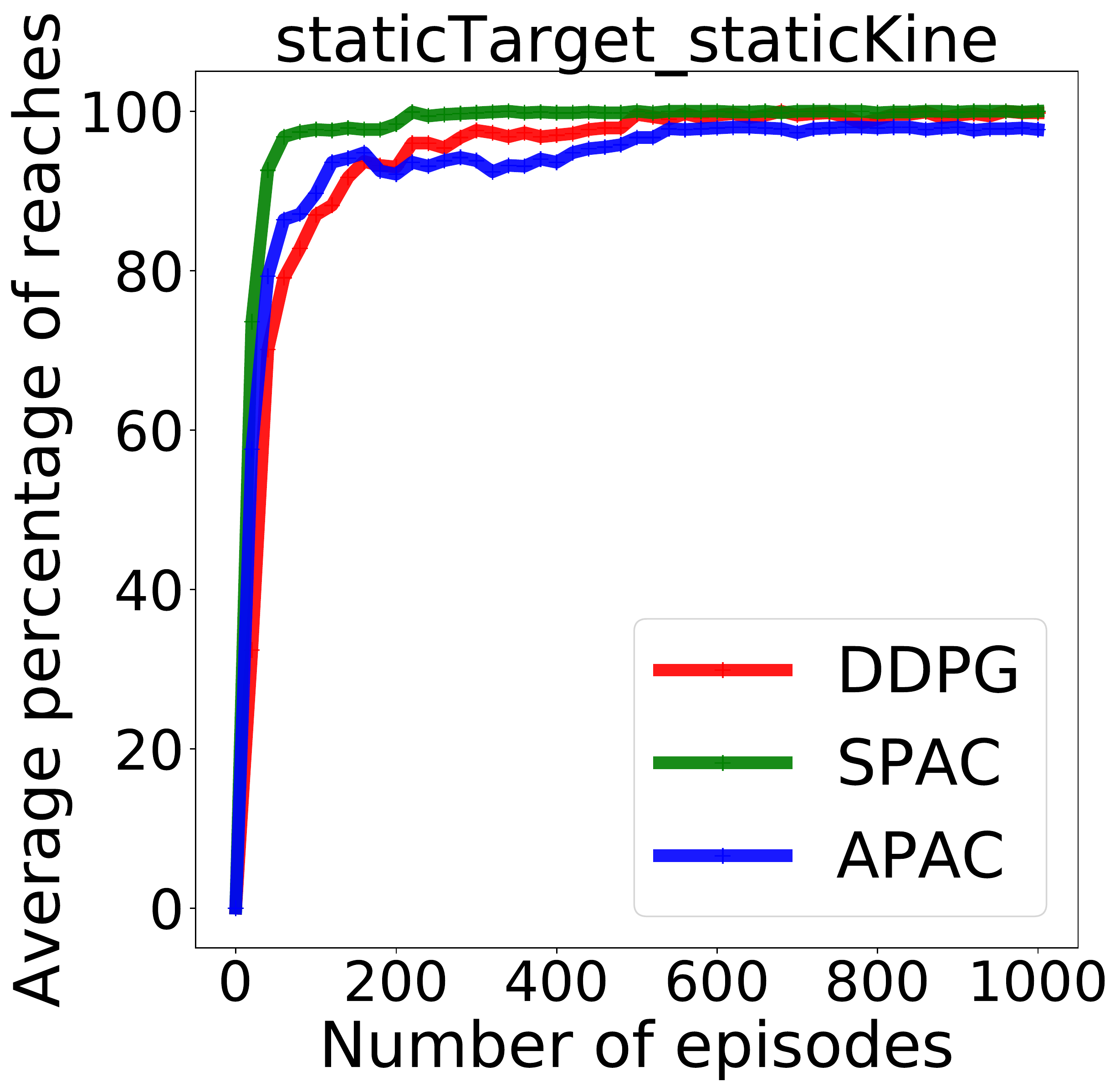}
\includegraphics[width=.24\textwidth]{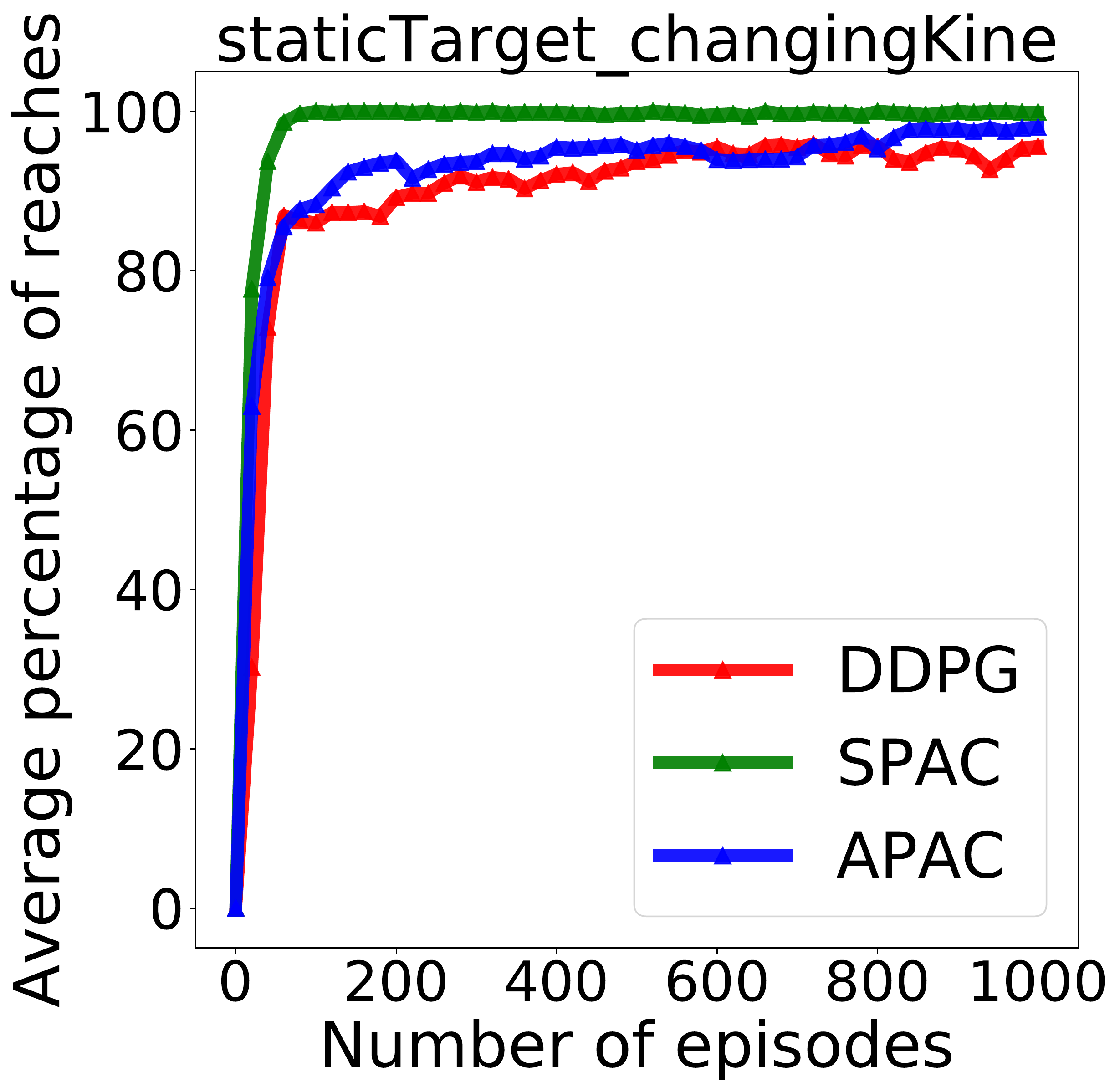}
\includegraphics[width=.24\textwidth]{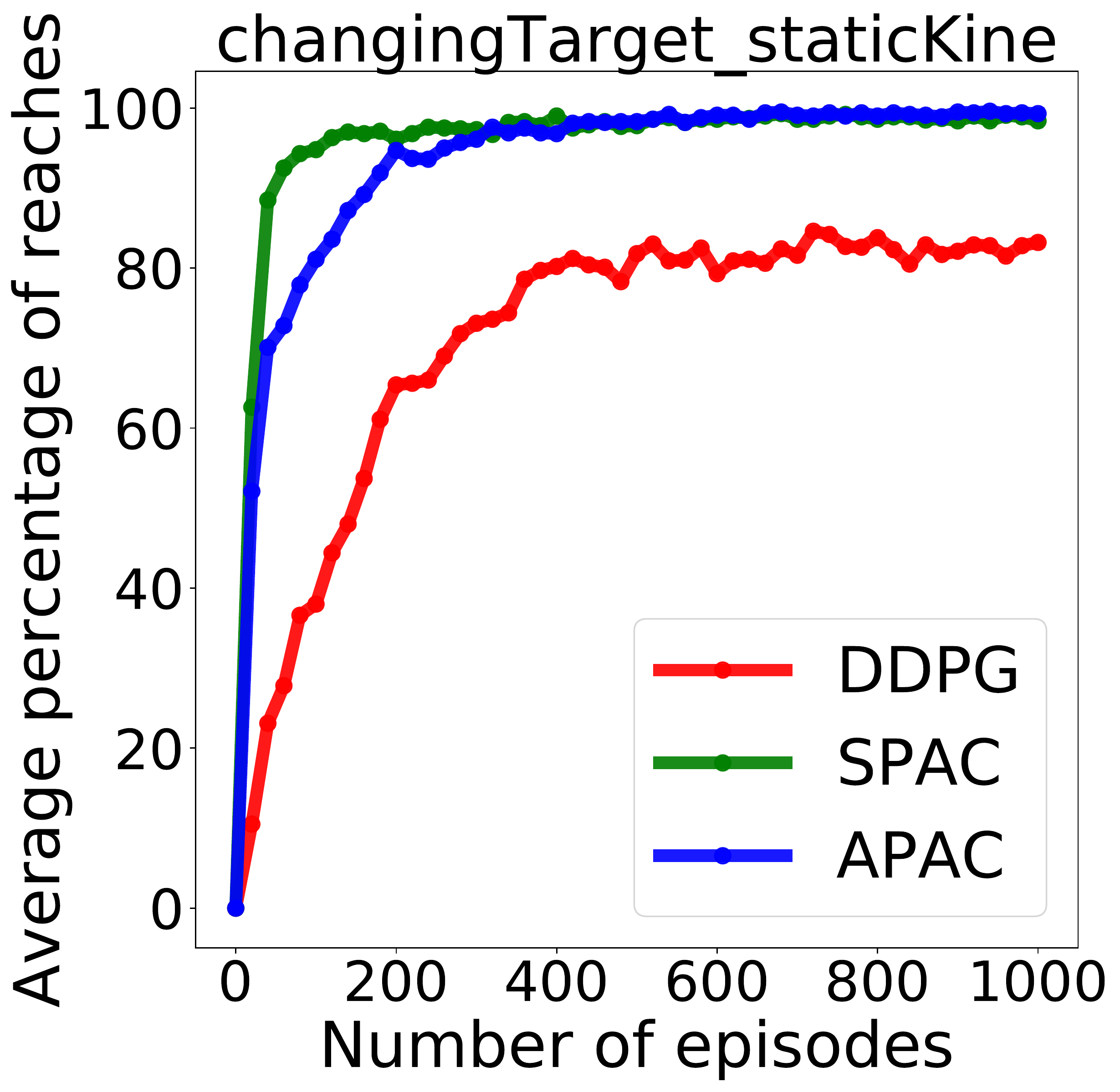}
\includegraphics[width=.24\textwidth]{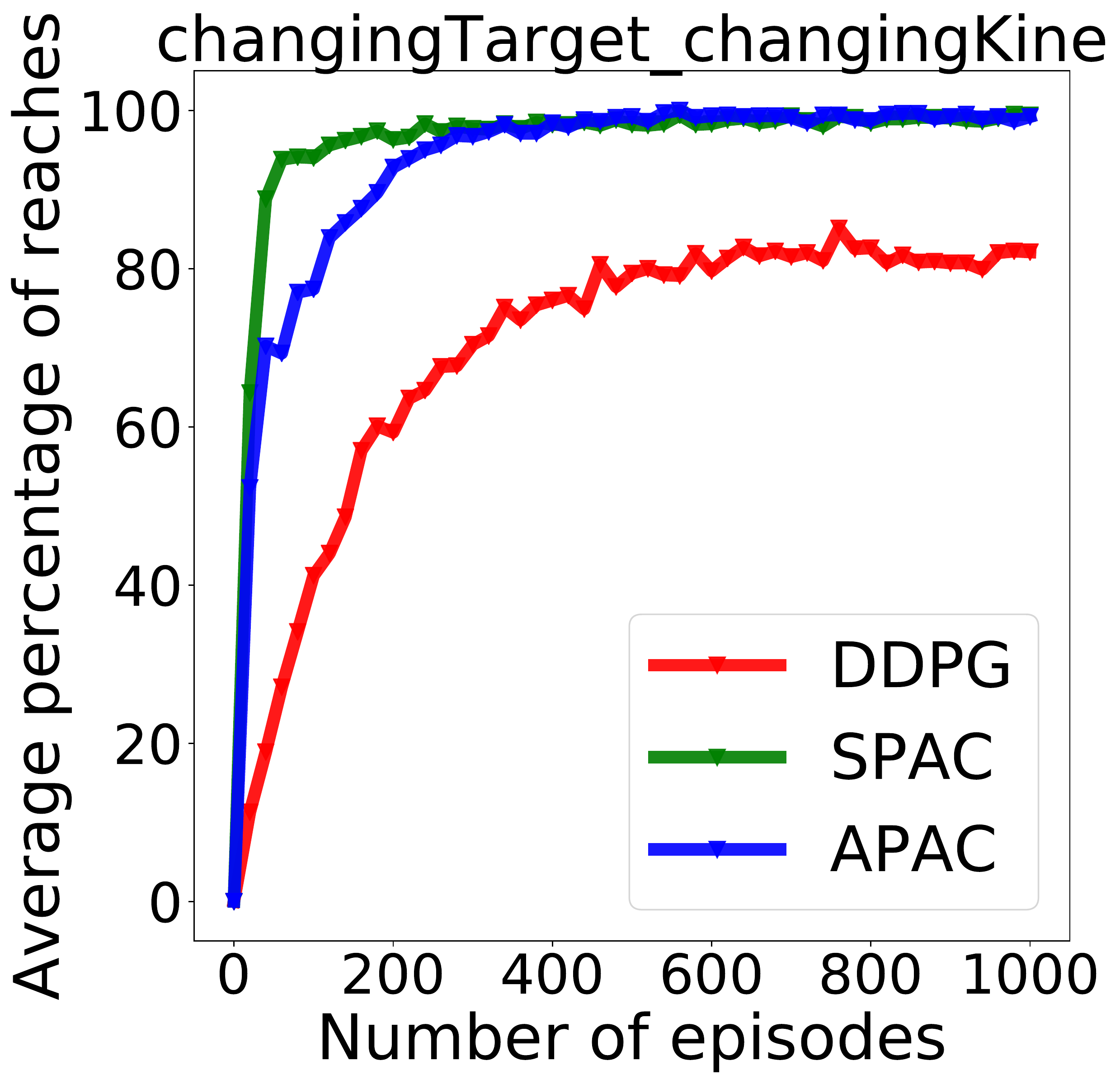}

\caption[comparison]{Comparison of the proportion of trials in which the target was successfully reached during training between DDPG, SPAC, and APAC. {\bf Top row} highlights differences in each respective model's performance across the Target/Kinematic experimental combinations. APAC that arbitrates between planning and habits can resist to changes in the target position and kinematic changes, while DDPG is not flexible under either changing condition.  {\bf Bottom row} plots highlight differences between models within each experimental combination of target position and kinematics. Performance of DDPG (pure habits) drops when target location is changing, while this has no effect on SPAC (exclusive planning) and APAC (arbitrated). }
   \label{comparison_modelCond}
\end{figure*}

\begin{figure*}[ht]
 \centering
     \includegraphics[width=.24\textwidth]{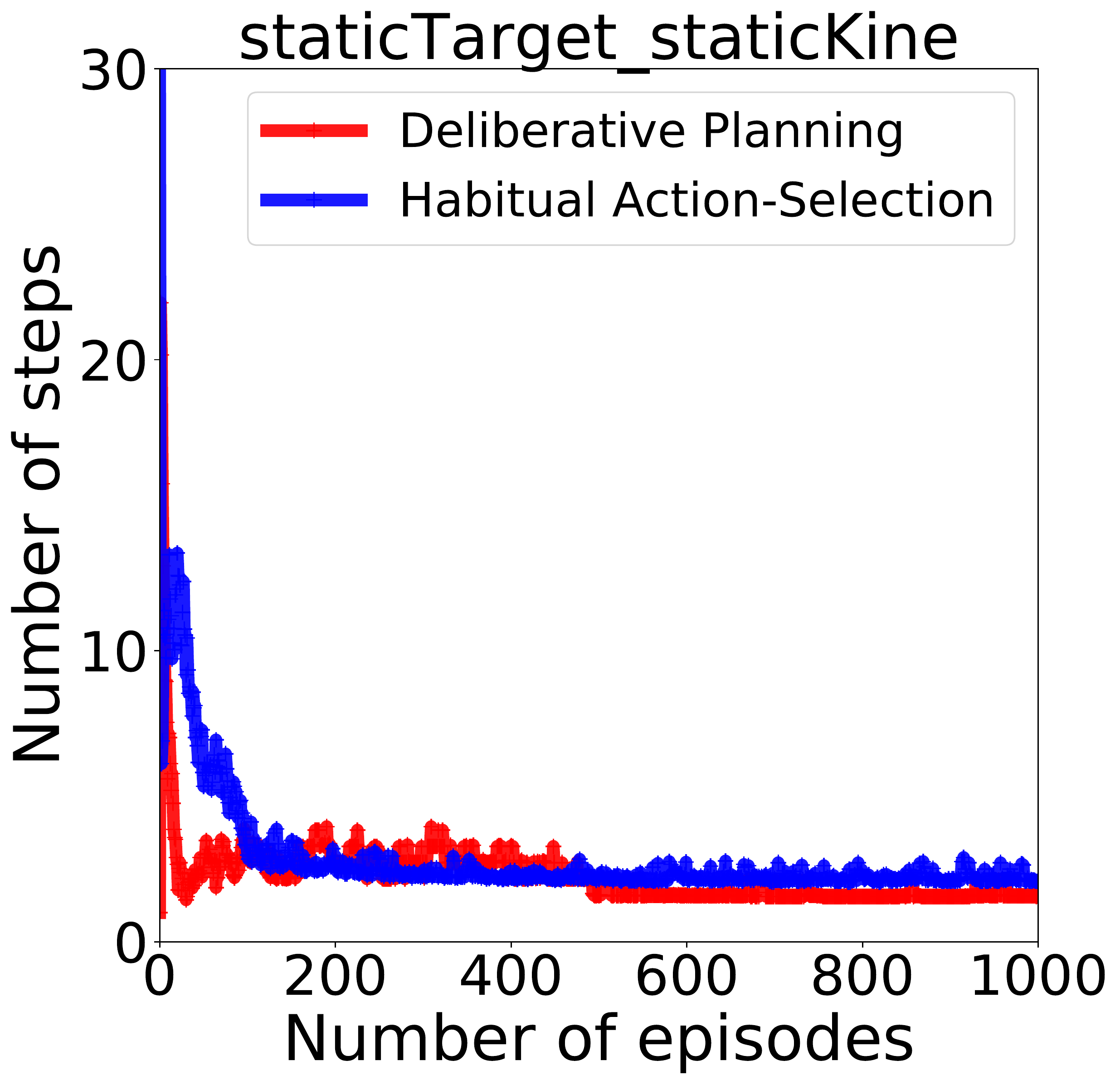}%
     \includegraphics[width=.24\textwidth]{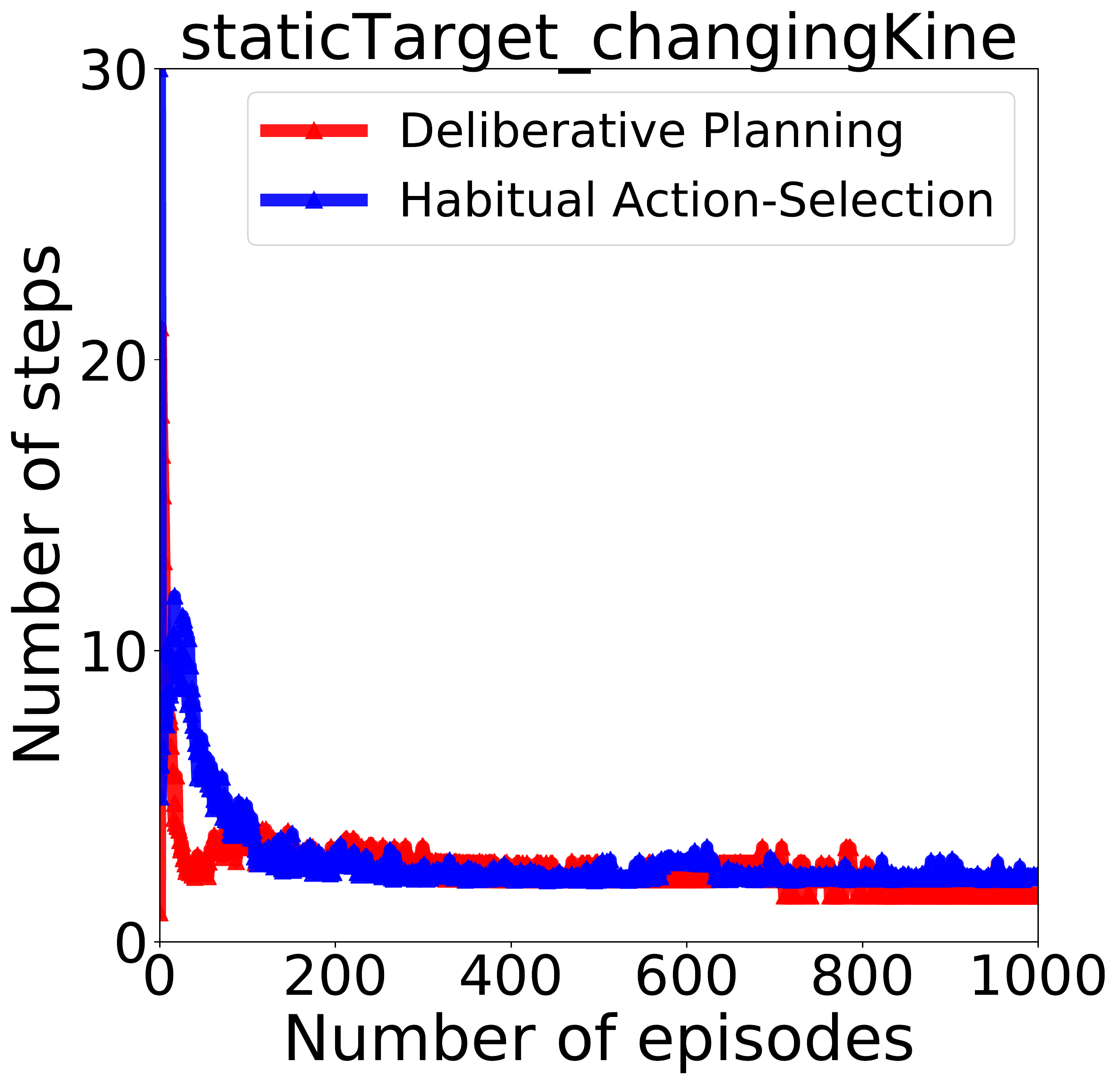}%
 	 \includegraphics[width=.24\textwidth]{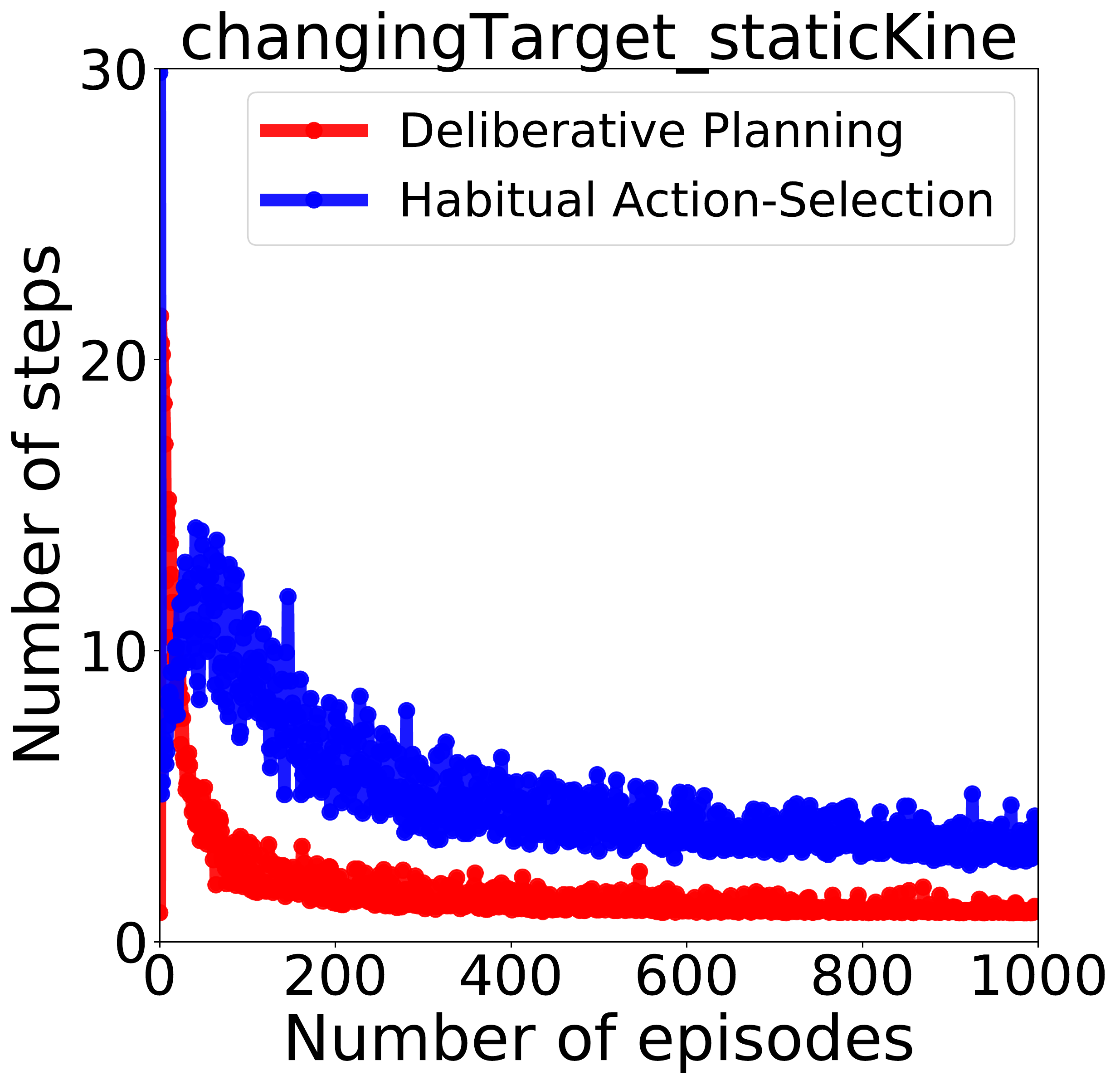}%
     \includegraphics[width=.24\textwidth]{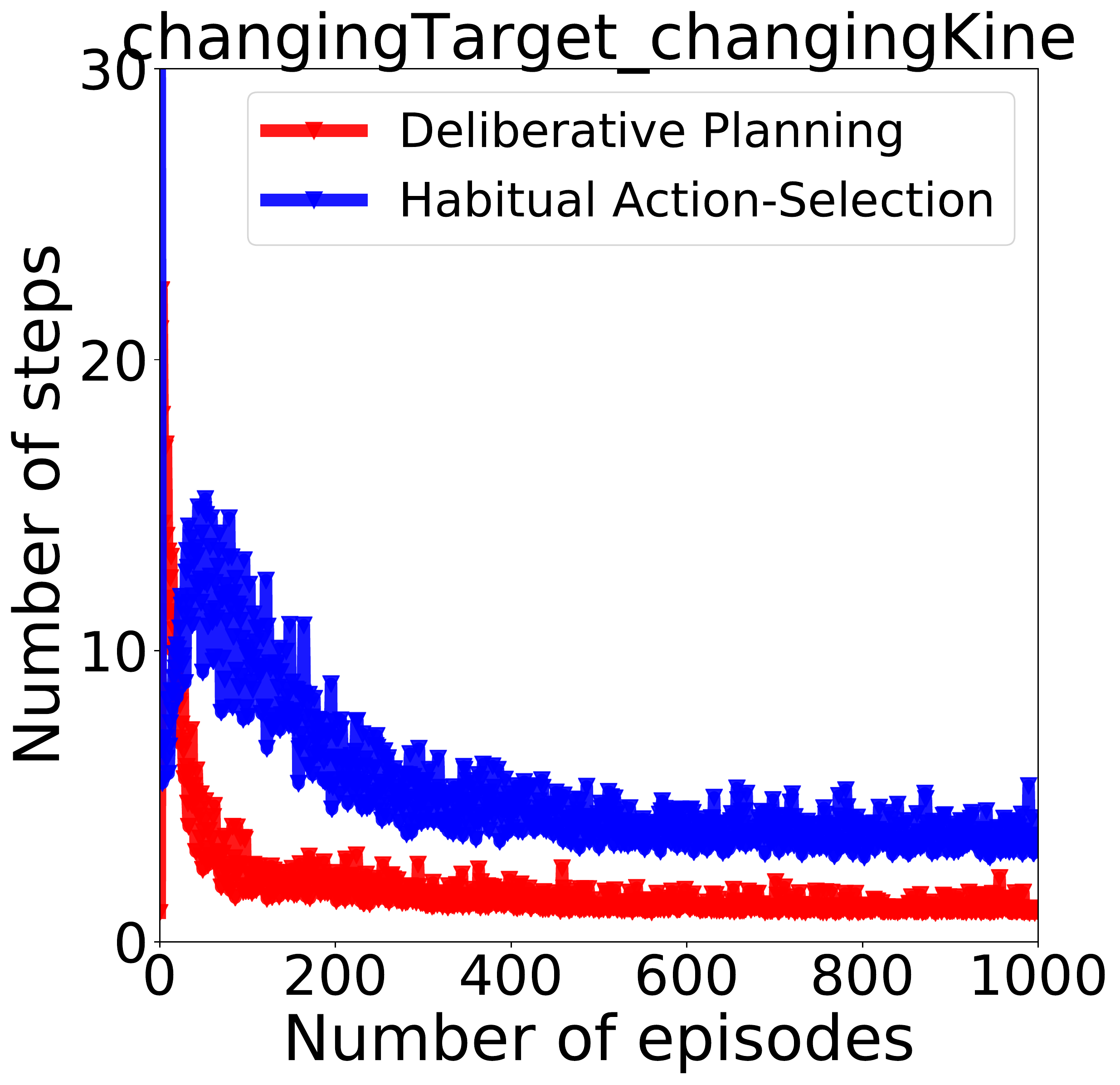}%
     \caption[APAC_arbitration]{Arbitration between habitual and planning by the APAC: For each condition, the blue line indicates the average number of actions which are selected by the habitual controller (i.e. the actor), while the red line demonstrates the average number of actions selected by the planning controller (i.e. inverse model). Results illustrate that planning controller is used early in the training, while later agent tends to use the habits more. }%
   \label{APAC_arbitration}%
\end{figure*}

Figure \ref{comparison_modelCond} illustrate the percentage of trials that reach the target within 30 action steps during training under different conditions for 1000 episodes each. The pure deliberate planning model SPAC reaches almost near perfect performance very fast after 100 episodes. Furthermore, SPAC's performance is very robust under different conditions and neither changes in reward function nor in kinematics effects the performance of SPAC very much. DDPG learns to reach almost 100\% of the targets only under static target/static kinematics condition. Performance of DDPG drops slightly under changing kinematics compared to static kinematics; however, its performance drops dramatically (about 20\%) under changing target conditions. This is of course expected as habits become invalid solutions under changing environments. Our point here is that APAC can reach almost all of the targets both under static and changing targets as good as SPAC, although it tends to use more habits than planning after a few trials of learning(see Figure \ref{APAC_arbitration}). The speed of learning in APAC is also very high and comparable to SPAC. In this sense it combines the benefits of DDPG and SPAC.

\begin{figure*}[!h]
\centering
\includegraphics[width=.7\textwidth]{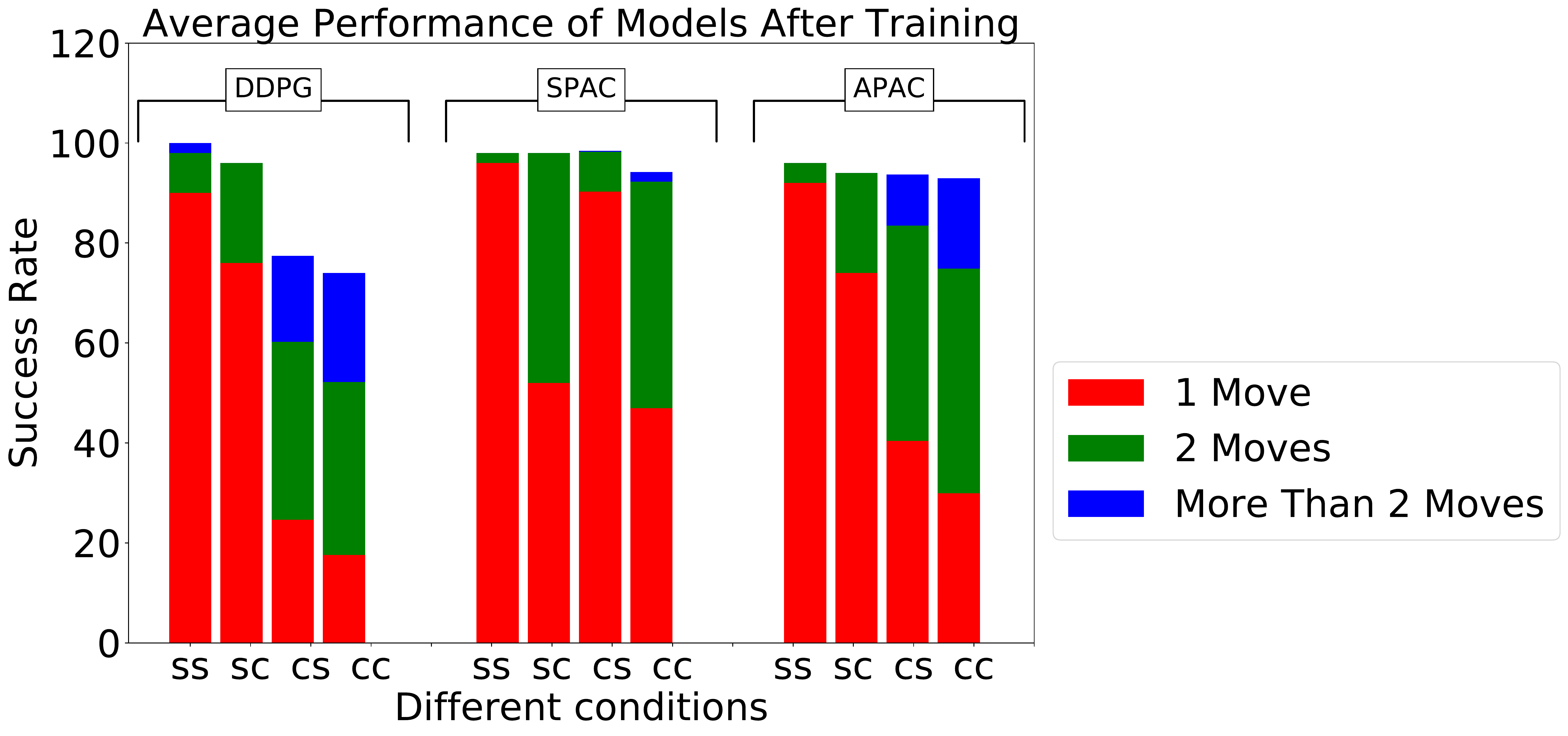}
\caption{Success rate to reach targets during testing by three models under different conditions, where 'ss', 'sc', 'cs', and 'cc' stand for static target/static kinematics, static target/changing kinematics, changing target/static kinematics, and changing target/ changing kinematics respectively. The plot also displays the average number of steps to reach 100 targets after training.}
   \label{testing_eachModel}
\end{figure*}

The above curves give an example of behaviour of the models during one learning trials. To study how these results generalize we tested the performance of all three models after learning over 50 different learning trials with random initial conditions for the networks. For the static target location we tested on the target location, that was randomly chosen for each learning trial. However, with the changing target location we decided to cover the possible target locations more systematically and chose a set target points on a regular grid in angle space. Figure \ref{testing_eachModel} displays average success rate over the 50 learning trials to reach these targets. All three models under the static target/static kinematics condition reach 100\% of the targets.
DDPG and APAC have slightly less success under static target/changing kinematics, while SPAC stays flexible under this condition. The major difference between DDPG and APAC become clear under changing target conditions, where DDPG's performance drops dramatically, while APAC obtains very good performance. SPAC is still very flexible to reach targets under changing target conditions. Overall it is remarkable how close APAC stays to the overall performance of SPAC in a situation where deliberative planning is the better choice.   

The overall success rate does show the entire range of the solutions. We thus included the individual performances in terms of the average number of steps to reach the target. As can be seen, APAC needs to take sometimes more corrective steps to reach the target while an exclusive planning system can optimize the number of steps. This is interesting as this allows for different strategies in solving the task, that of relying somewhat on habitual control when the cost of the movement initiation might be small versus more deliberate planning when the number of action steps might matter. This can explain a form of speed-accuracy trade-off.


\begin{figure*}[ht]
 \centering
	\includegraphics[width=.24\textwidth]{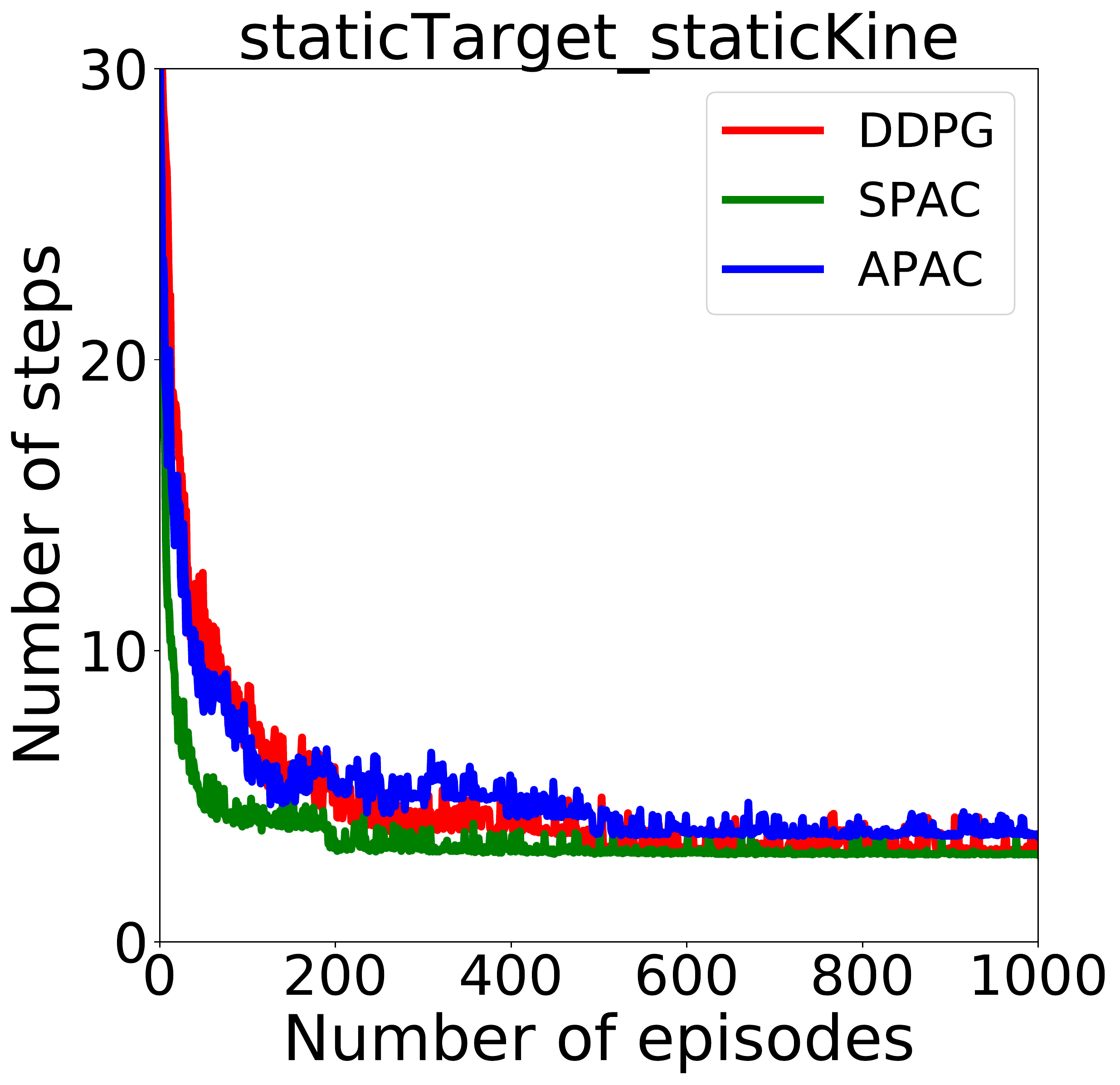}%
	\includegraphics[width=.24\textwidth]{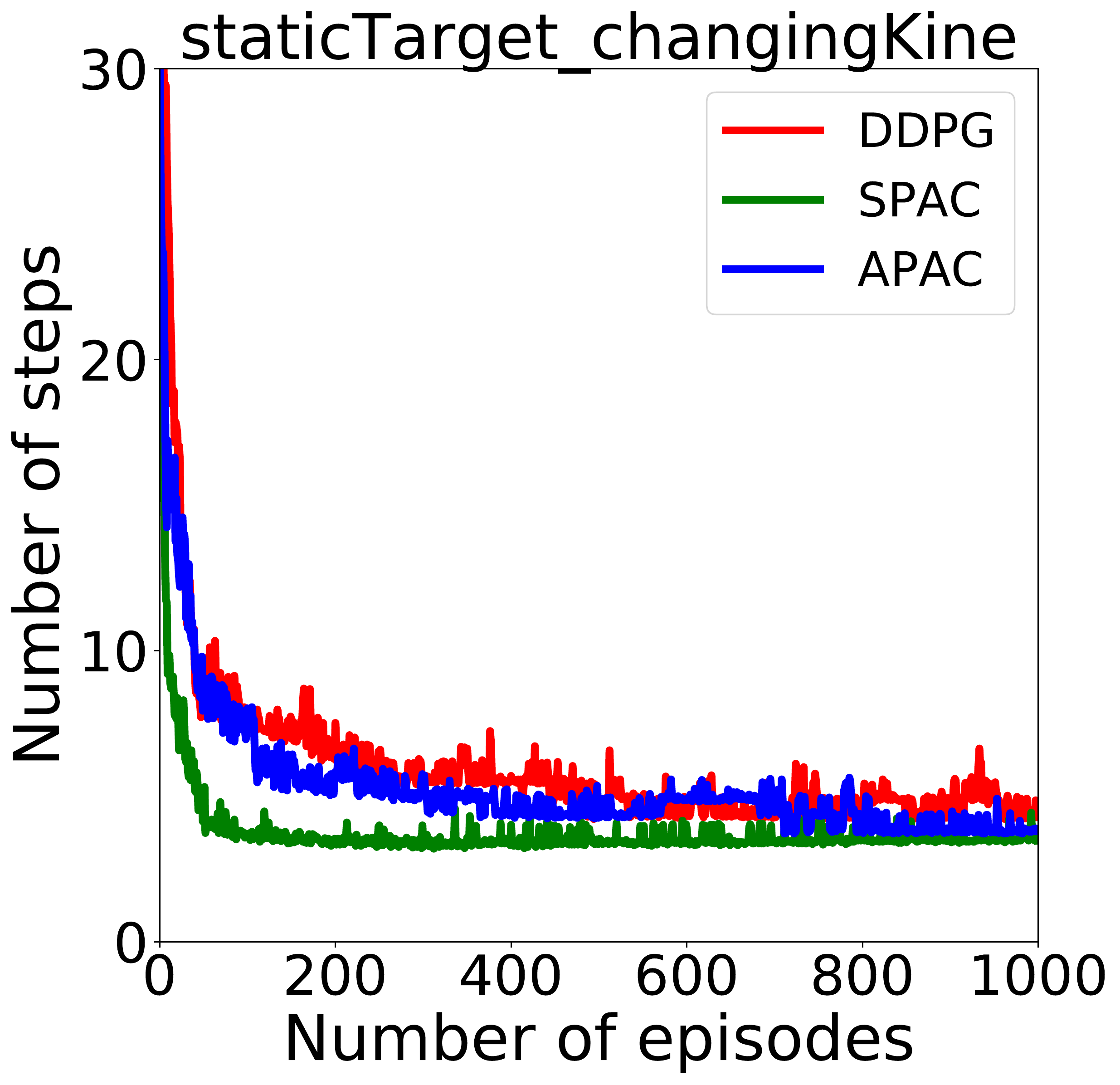}%
 	 \includegraphics[width=.24\textwidth]{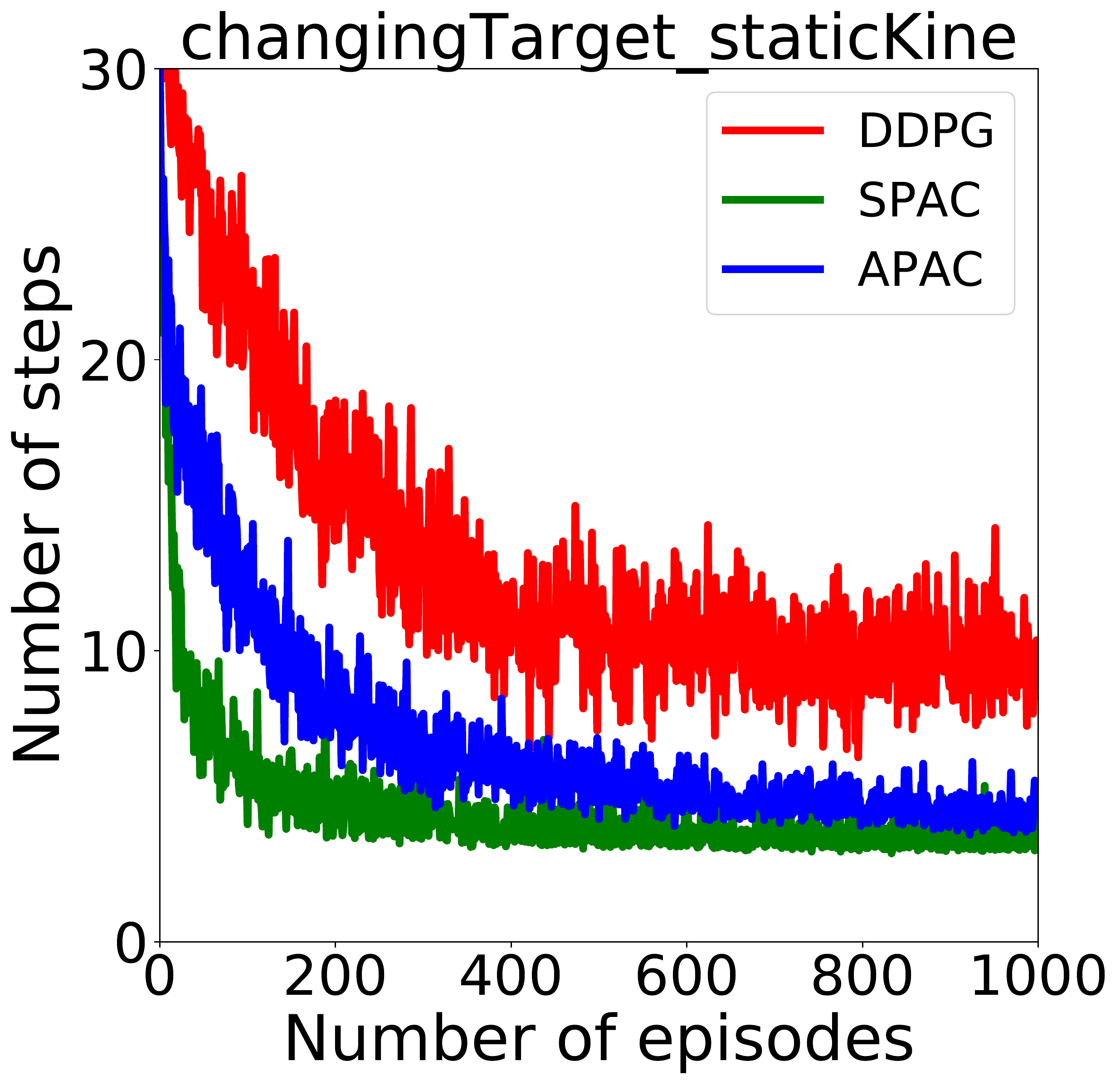}%
 	\includegraphics[width=.24\textwidth]{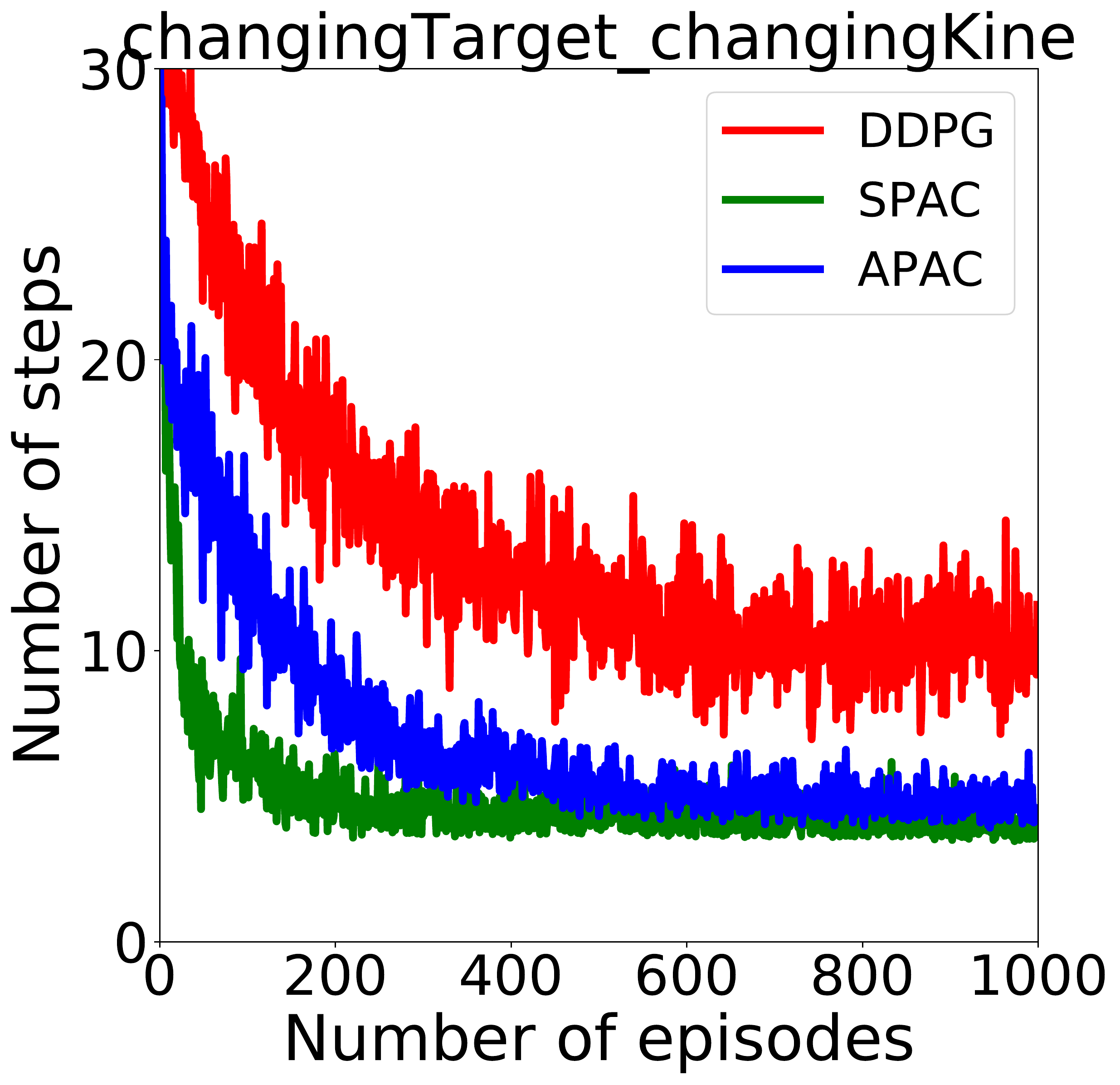}%
 	\\
	\includegraphics[width=.24\textwidth]{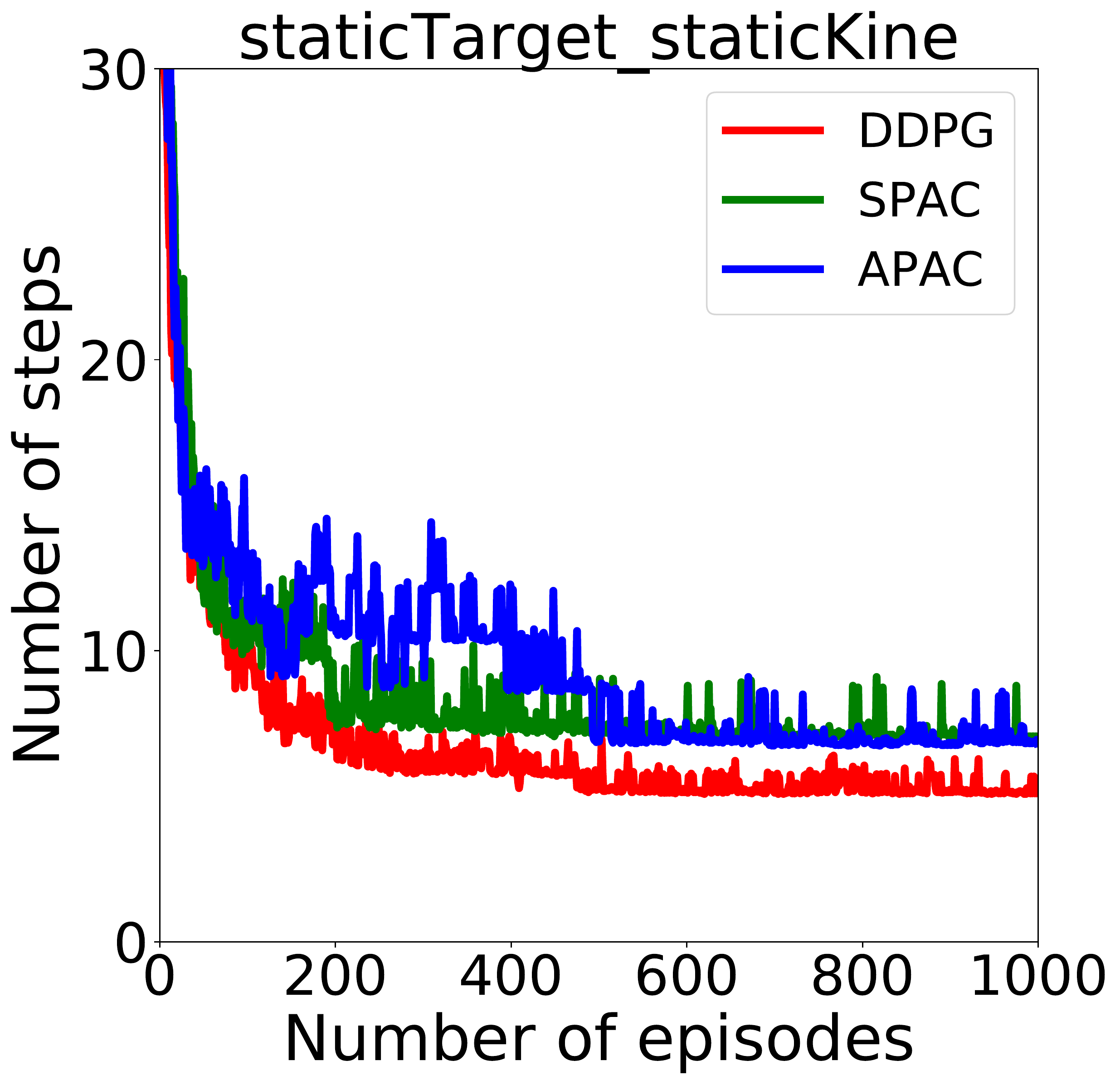}
     \includegraphics[width=.24\textwidth]{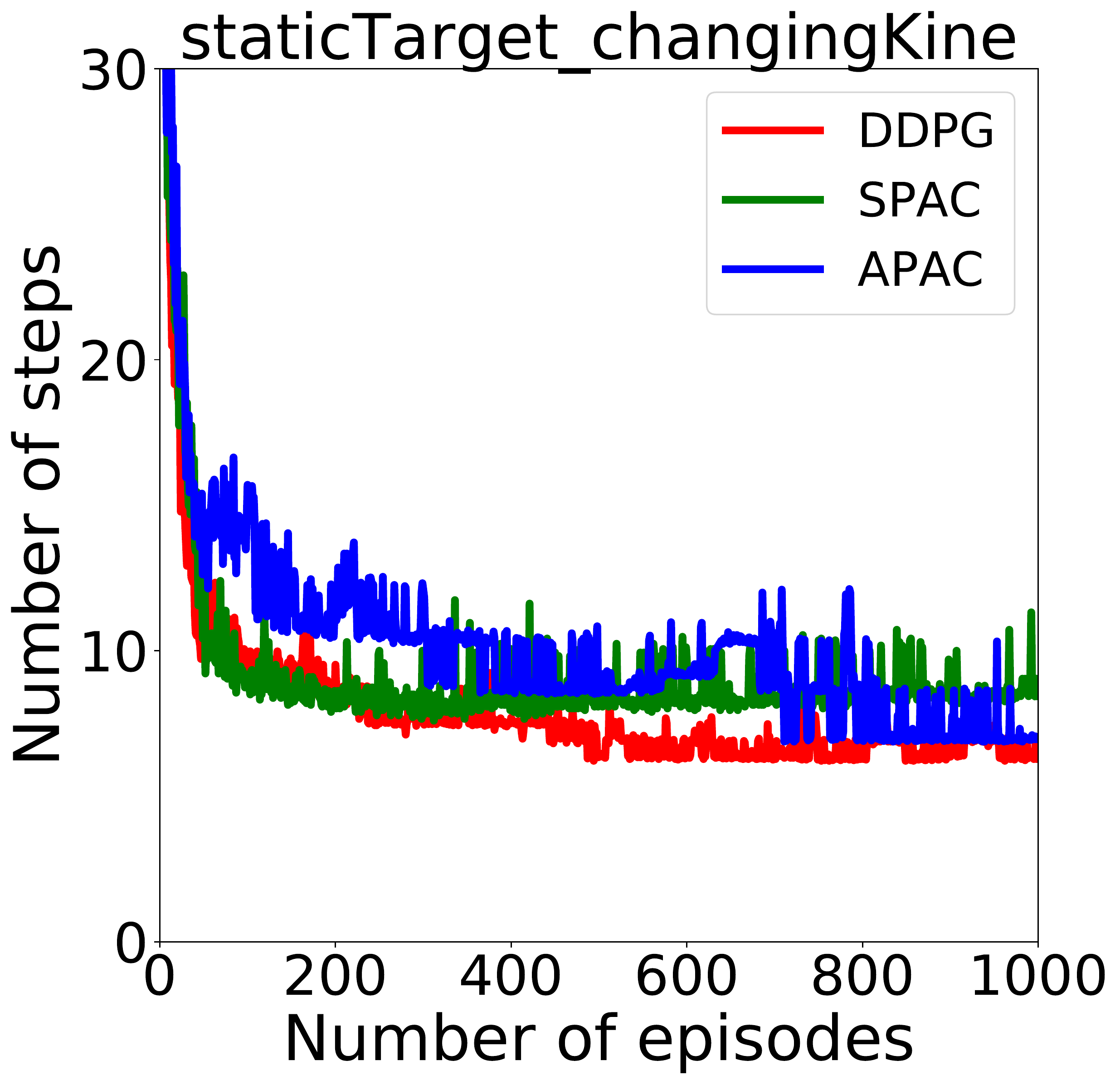}%
 	 \includegraphics[width=.24\textwidth]{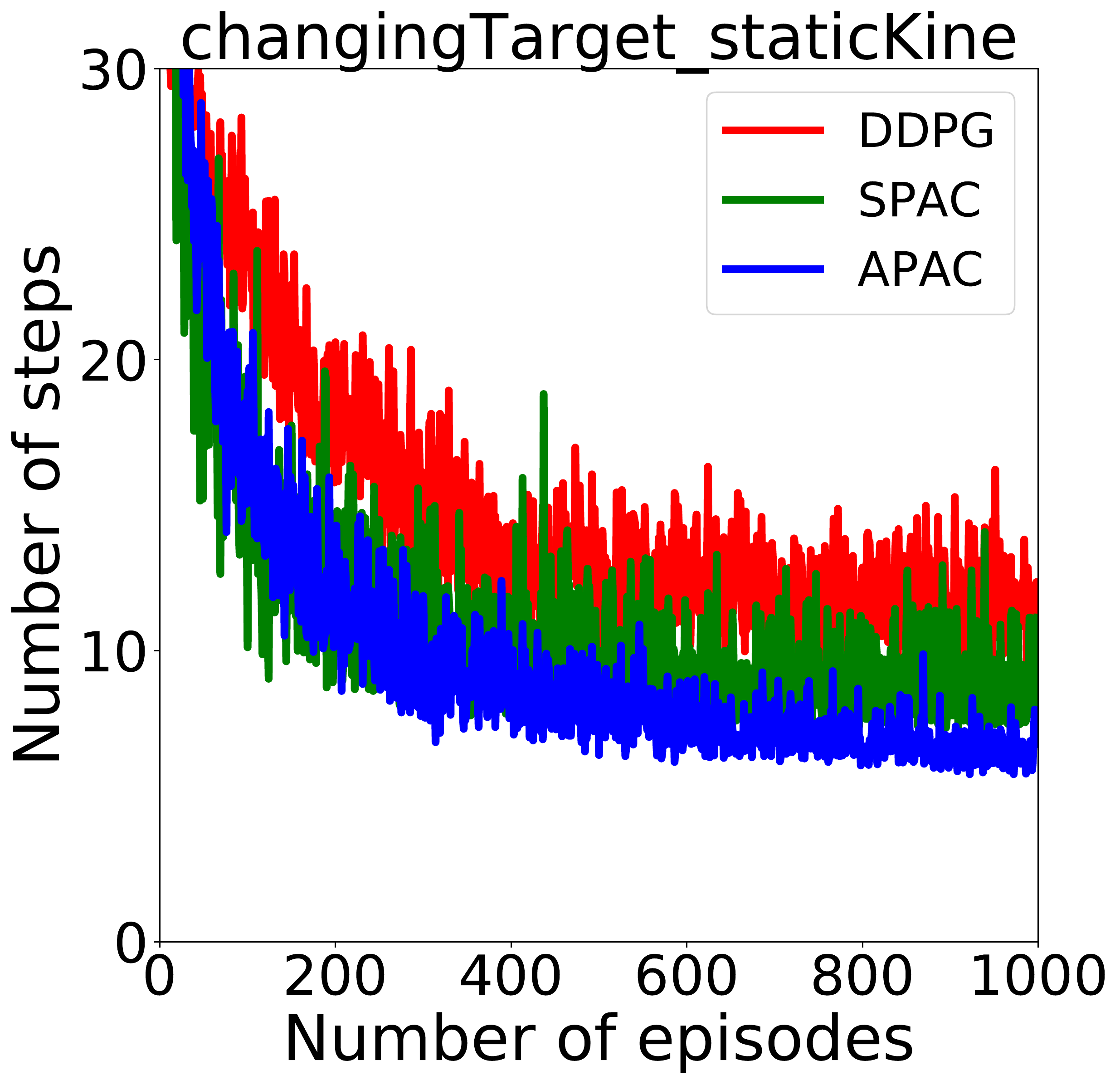}%
      \includegraphics[width=.24\textwidth]{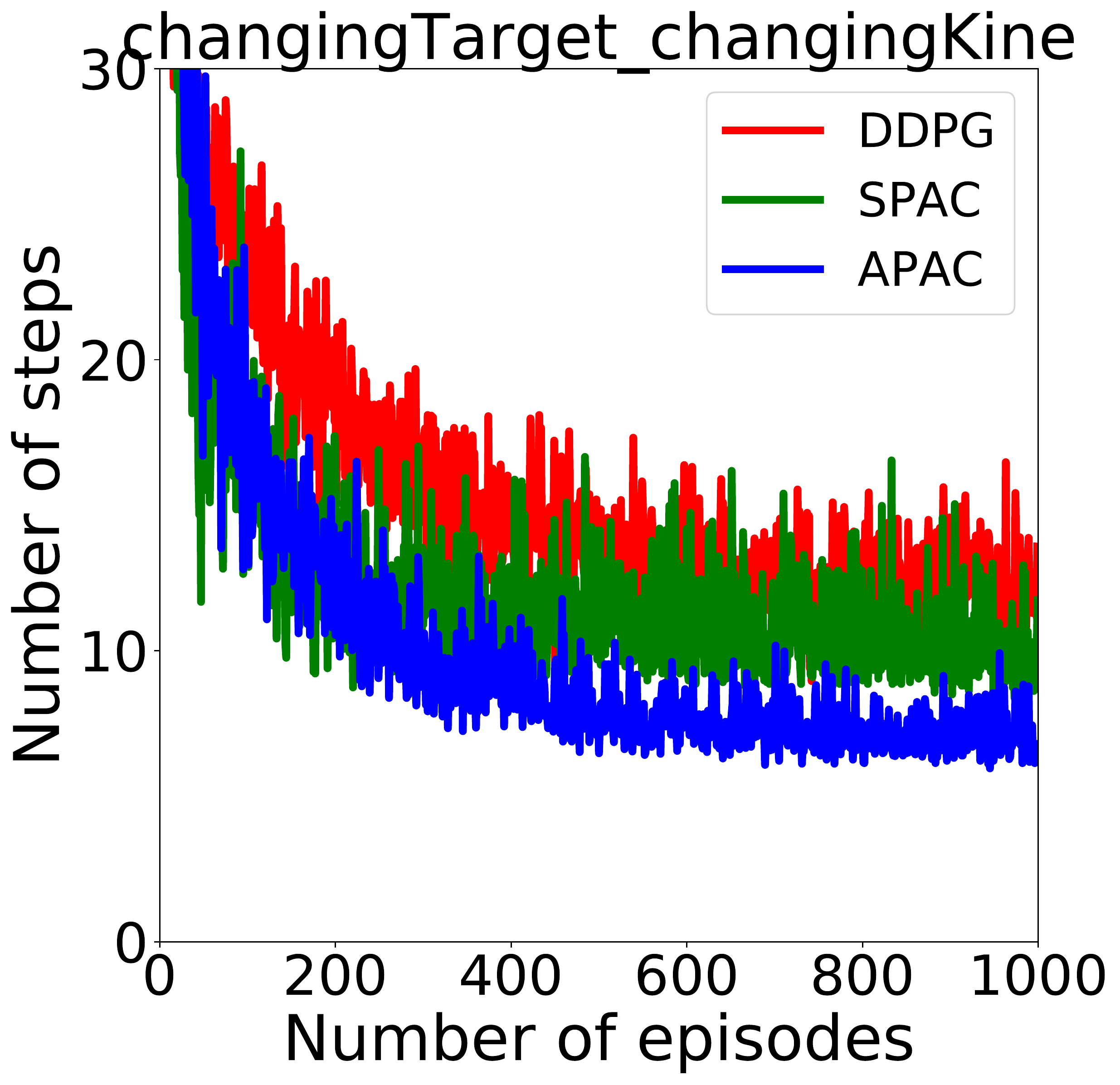}%

\caption{{\bf Top rpw:} Average action steps to complete the reaching task by each model under different conditions. {\bf Bottom row:} Average time steps to complete the reaching task at each episode under different conditions when a three-times higher cost for a planning control compared to the habitual control is taken into account.}%
   \label{APAC_arbitration_time}%
\end{figure*}

Figure \ref{APAC_arbitration} illustrates how APAC gradually shifts from a planning to a habitual control approach with increasing experience. After around 300 episode, more than 80\% of APAC actions were taken from the habitual controller (\ref{APAC_arbitration}). Of course, SPAC uses planning control throughout the entirety of the task, so no commensurate figure was generated for it. 

Since a habitual system should be faster than deliberate planning, this figure also illustrates that APAC would be less time-consuming than the SPAC at the same task and under the same condition. To show average time consumption by each model under different conditions, we assumed that each action selected by the deliberative planning takes three times longer than an action selected by the habitual controller. The number here is arbitrary and only chosen to show the general effect. The top row of plots in Figure \ref{APAC_arbitration_time} show the average number of action steps that each of the three models need to complete the task at each episode under different conditions.  These plots demonstrate that the number of steps are almost the same under static target/static kinematics  conditions. Under changing target conditions DDPG needs more steps to complete the task than APAC and SPAC. However, when including a higher cost for deliberative planning in the plots shown in the bottom row of the Figure \ref{APAC_arbitration_time}, the picture for the average number of time that is needed to complete the task changes. In this case, DDPG needs shorter time under static target conditions. However, under changing target conditions, APAC completes the task of reaching targets faster. DDPG takes longer to finish the task as it needs more corrective actions. 


\begin{figure*}[!t]
\centering
\includegraphics[width=.32\textwidth]{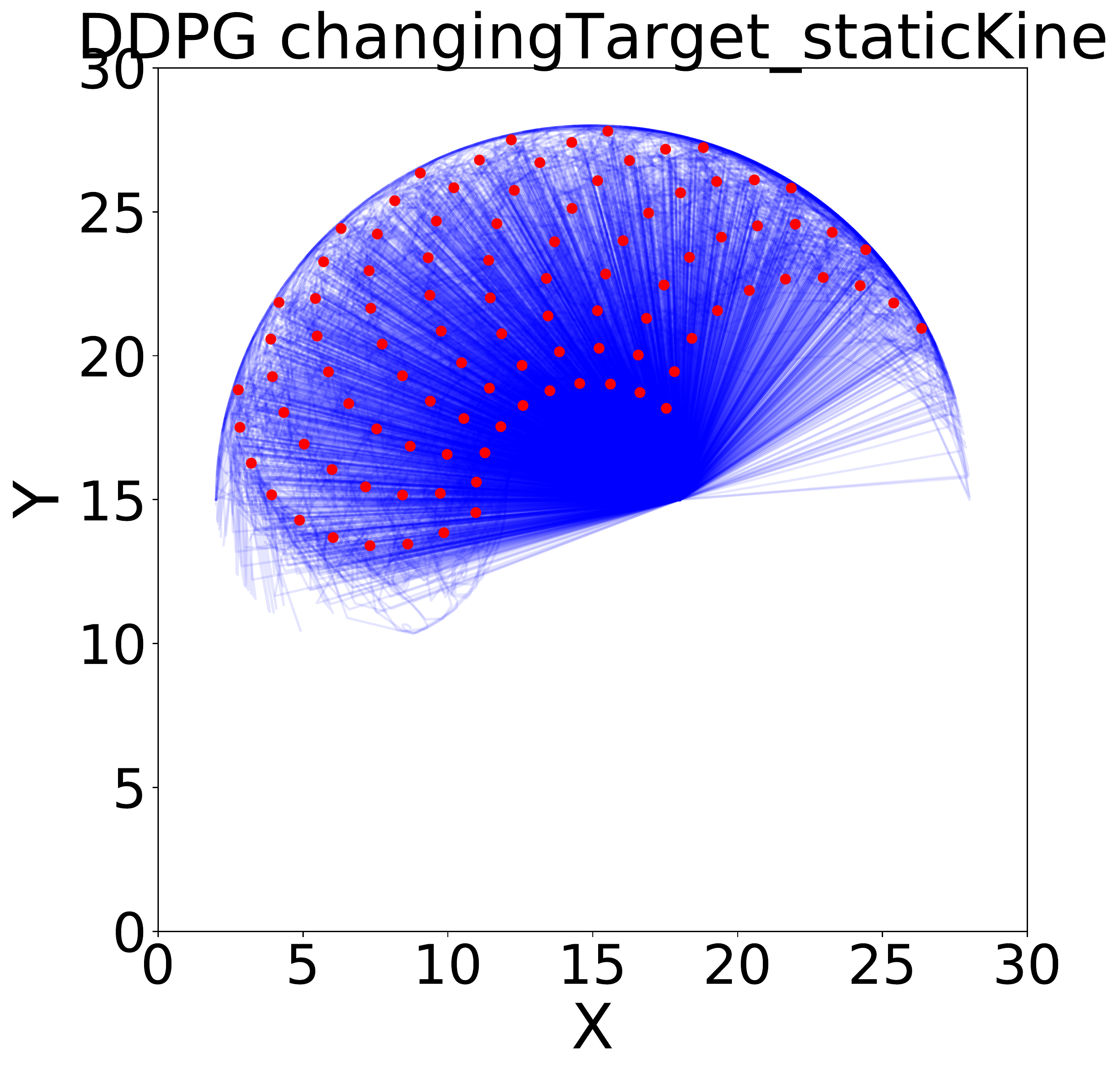}
\includegraphics[width=.32\textwidth]{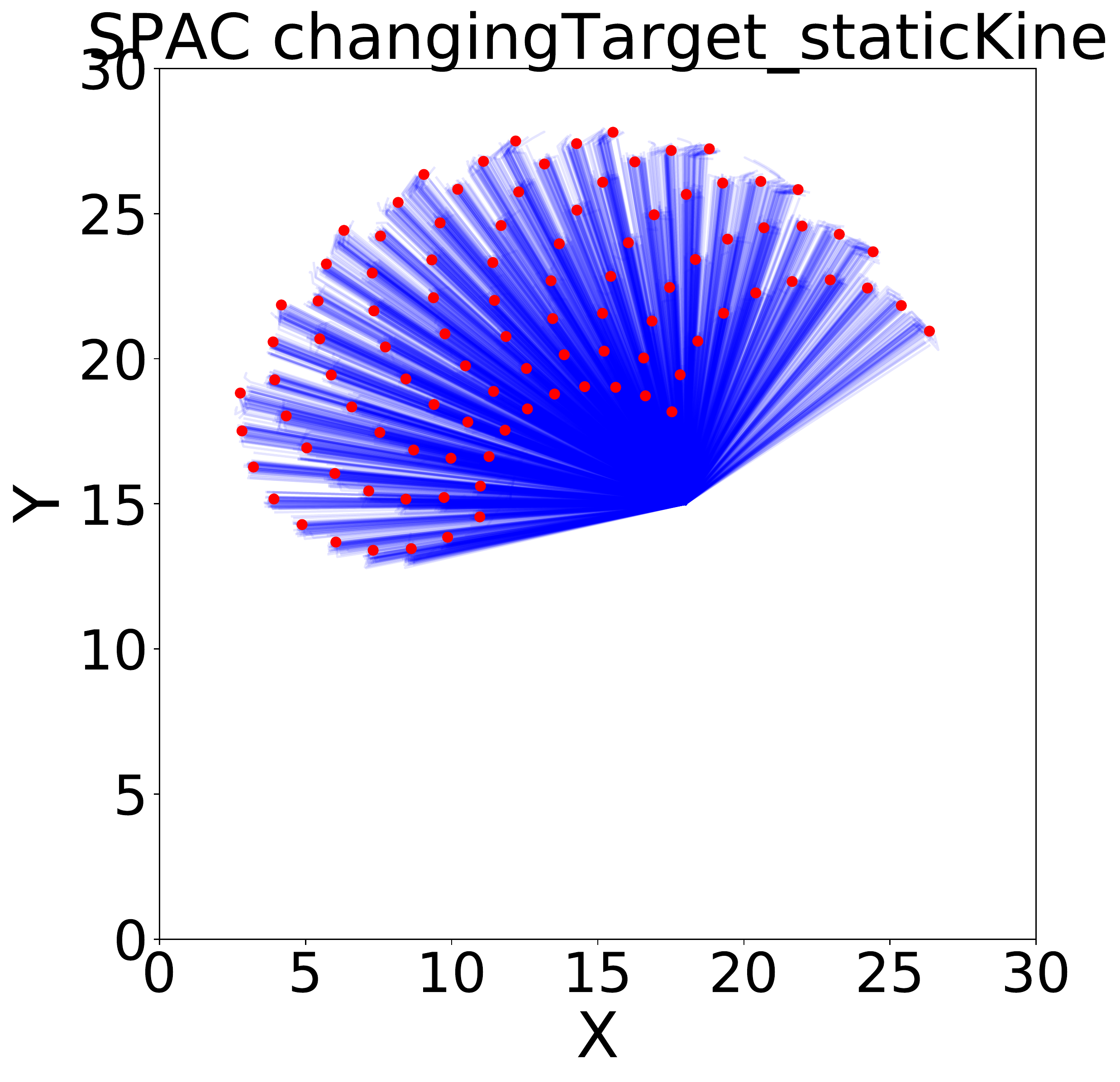}
\includegraphics[width=.32\textwidth]{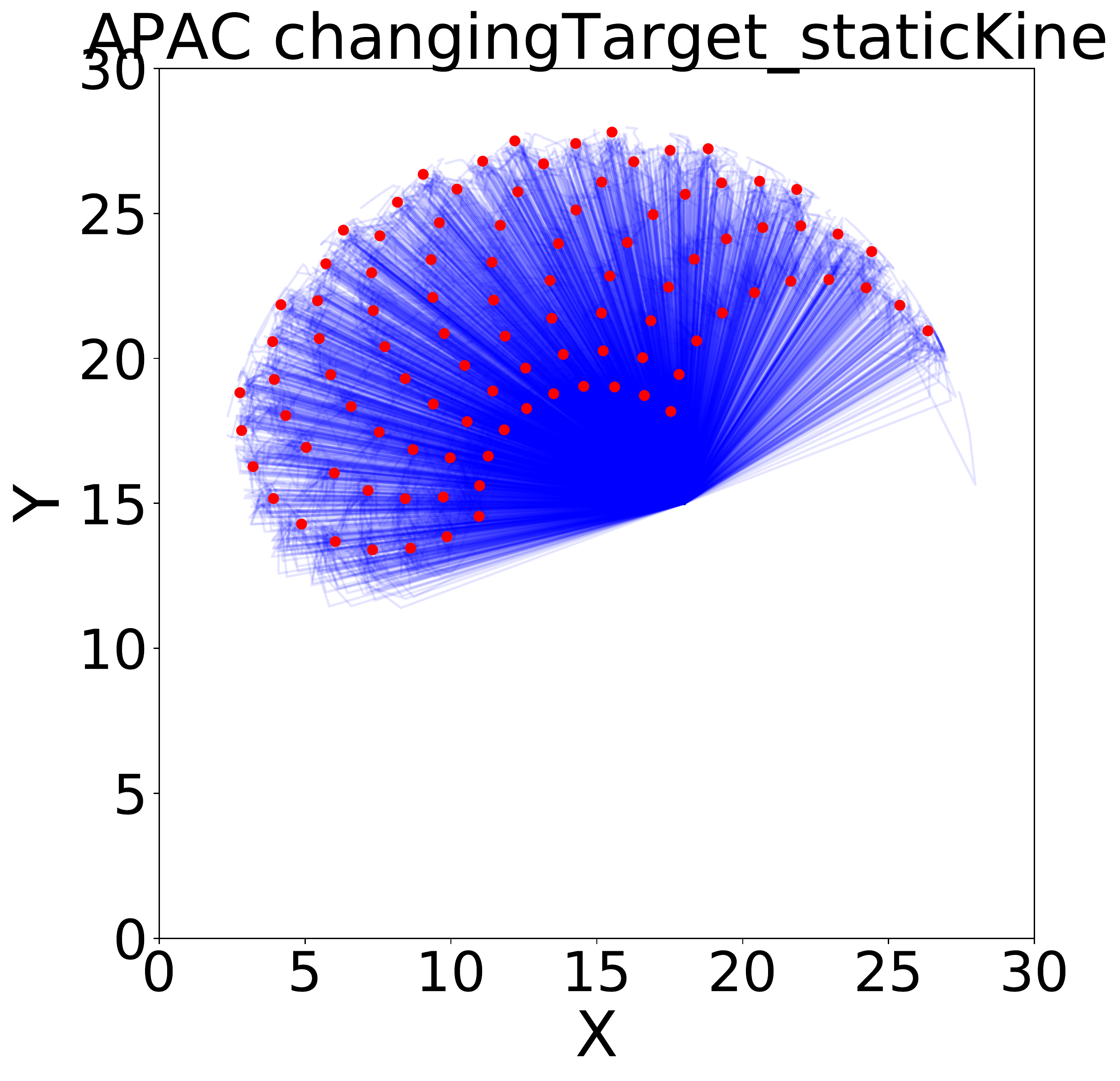}
\\
\includegraphics[width=.32\textwidth]{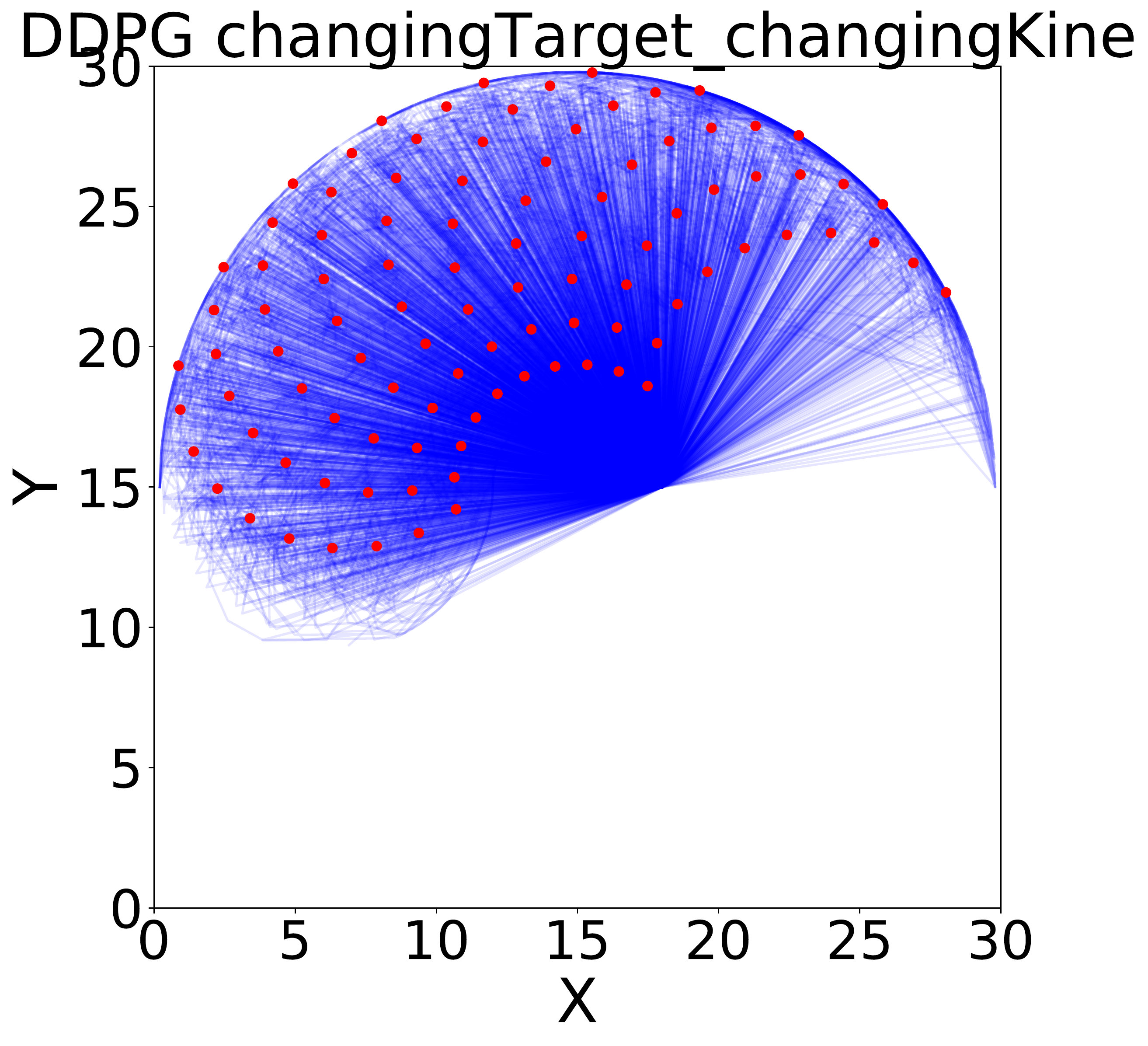}
\includegraphics[width=.32\textwidth]{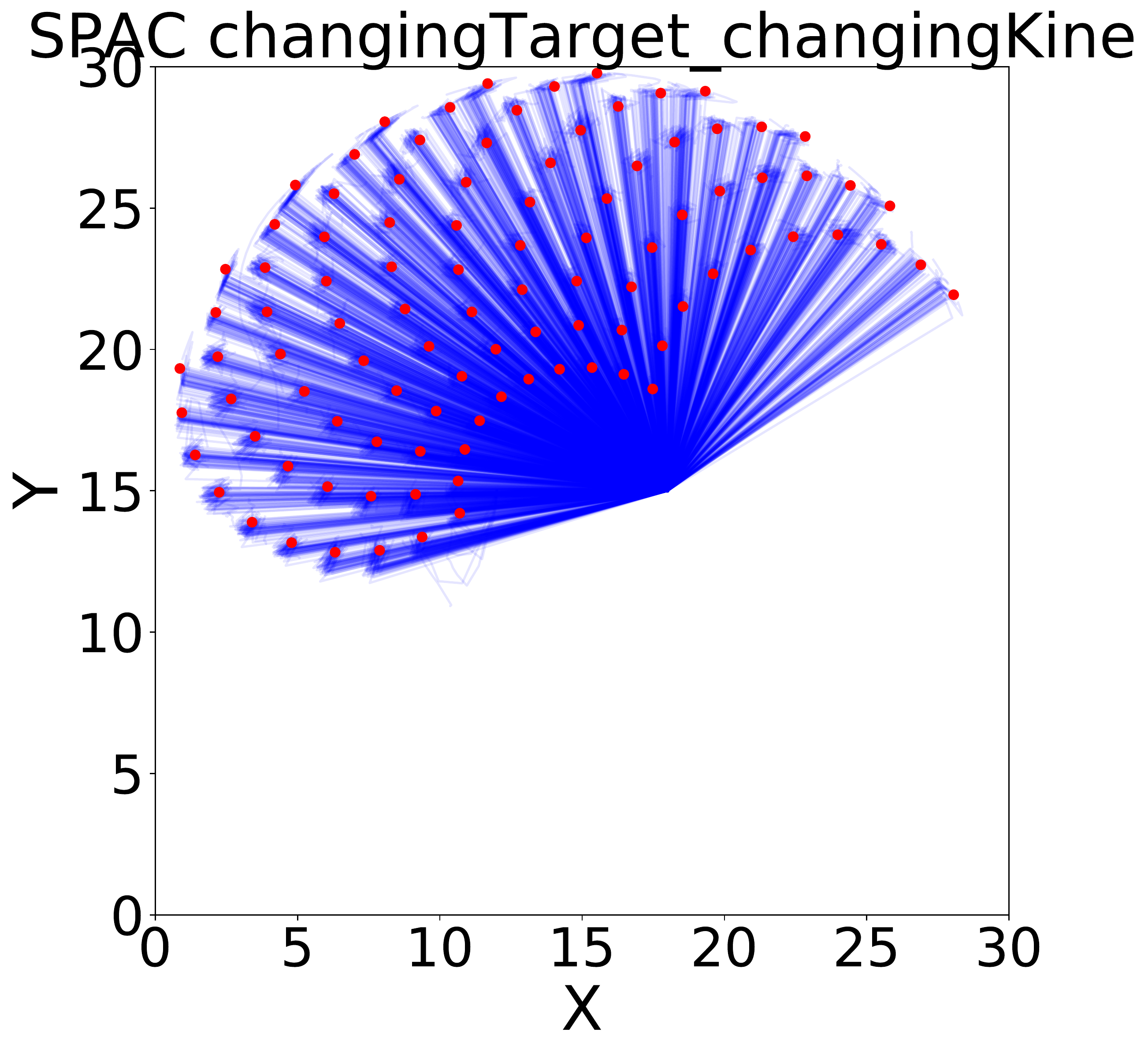}
\includegraphics[width=.32\textwidth]{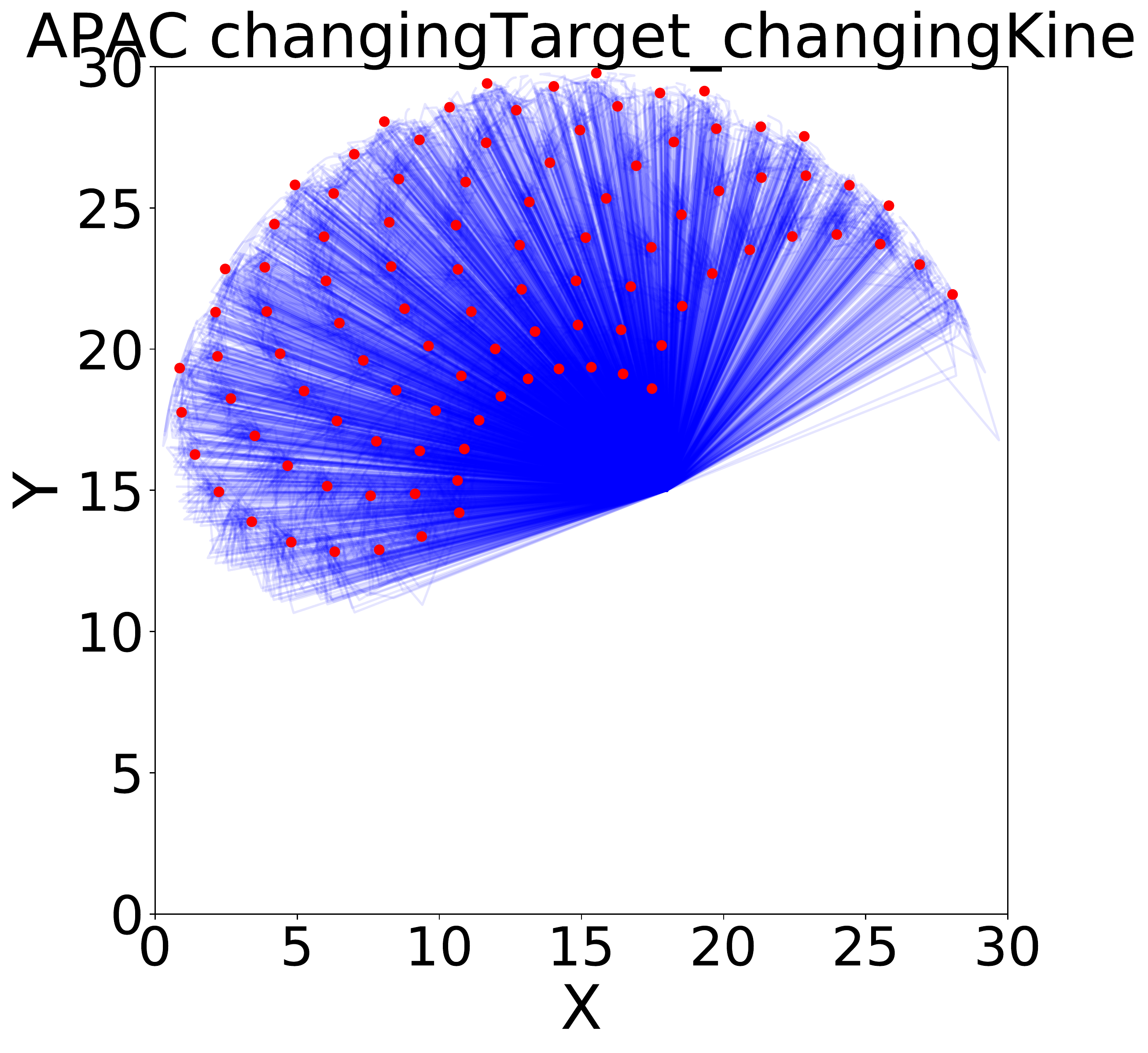}
\caption[reaching]{Reaching examples of three models under changing target conditions. Blue lines show position of the end-effector at each time step toward the target, while red dots are target locations during the testing phase. }
   \label{abc}
\end{figure*}
Figure \ref{abc} shows all 50 runs to reach 100 targets of a reaching test under changing target conditions, with (bottom row) and without (top row) changing kinematics, for all three models. The plots illustrate that DDPG has some difficulty reaching the locations at the edges of the possible target area due to non-linearity of the mapping between angles and locations. SPAC can learn the mapping function much better, and the quality of APAC is similar to SPAC. Interestingly, although APAC tends to use more habits than deliberate planning, this model can still reach many more targets than DDPG, almost as good as SPAC. 

\begin{figure*}[!t]
	\centering
	\includegraphics[width=0.3\textwidth]{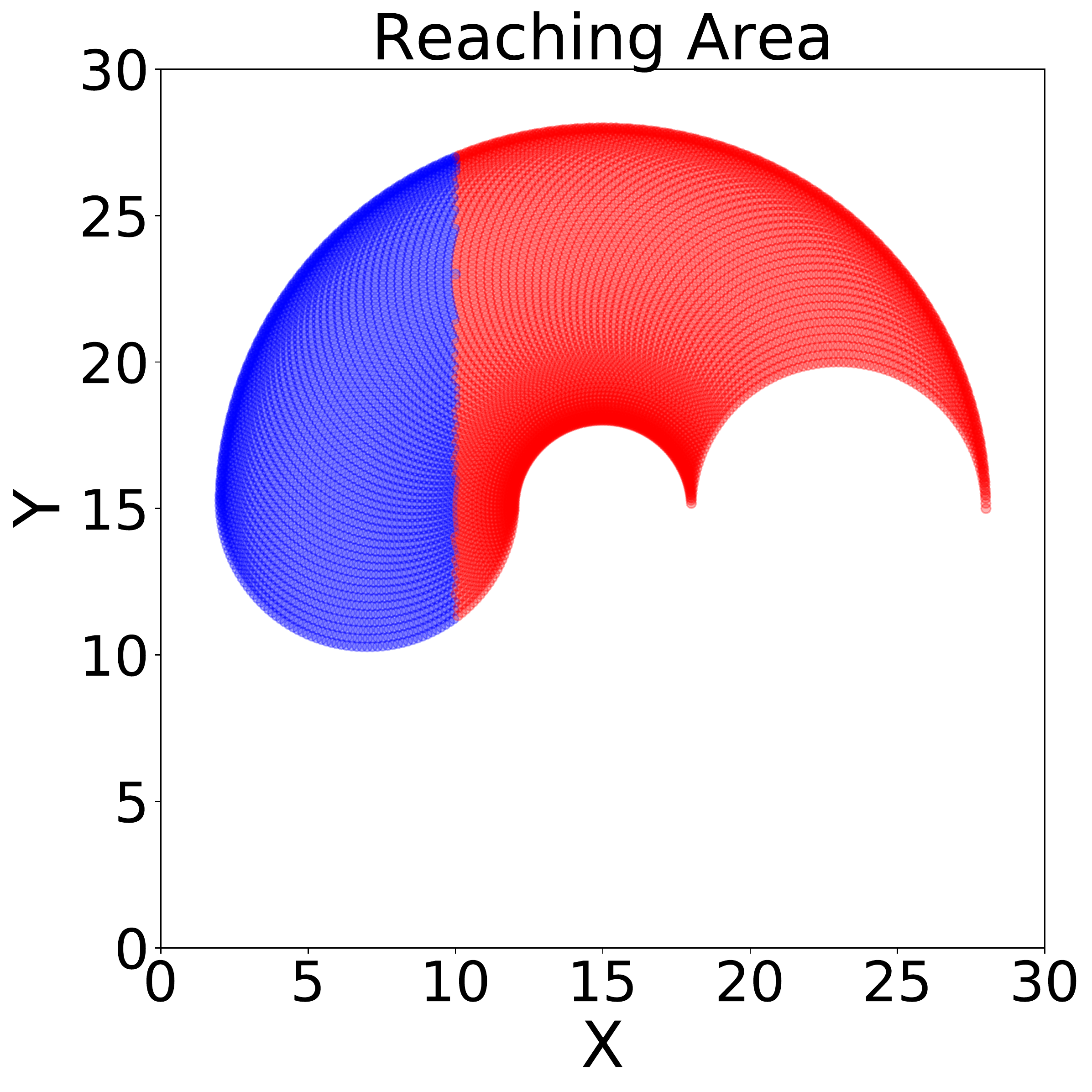}%
	\includegraphics[width=0.5\textwidth]{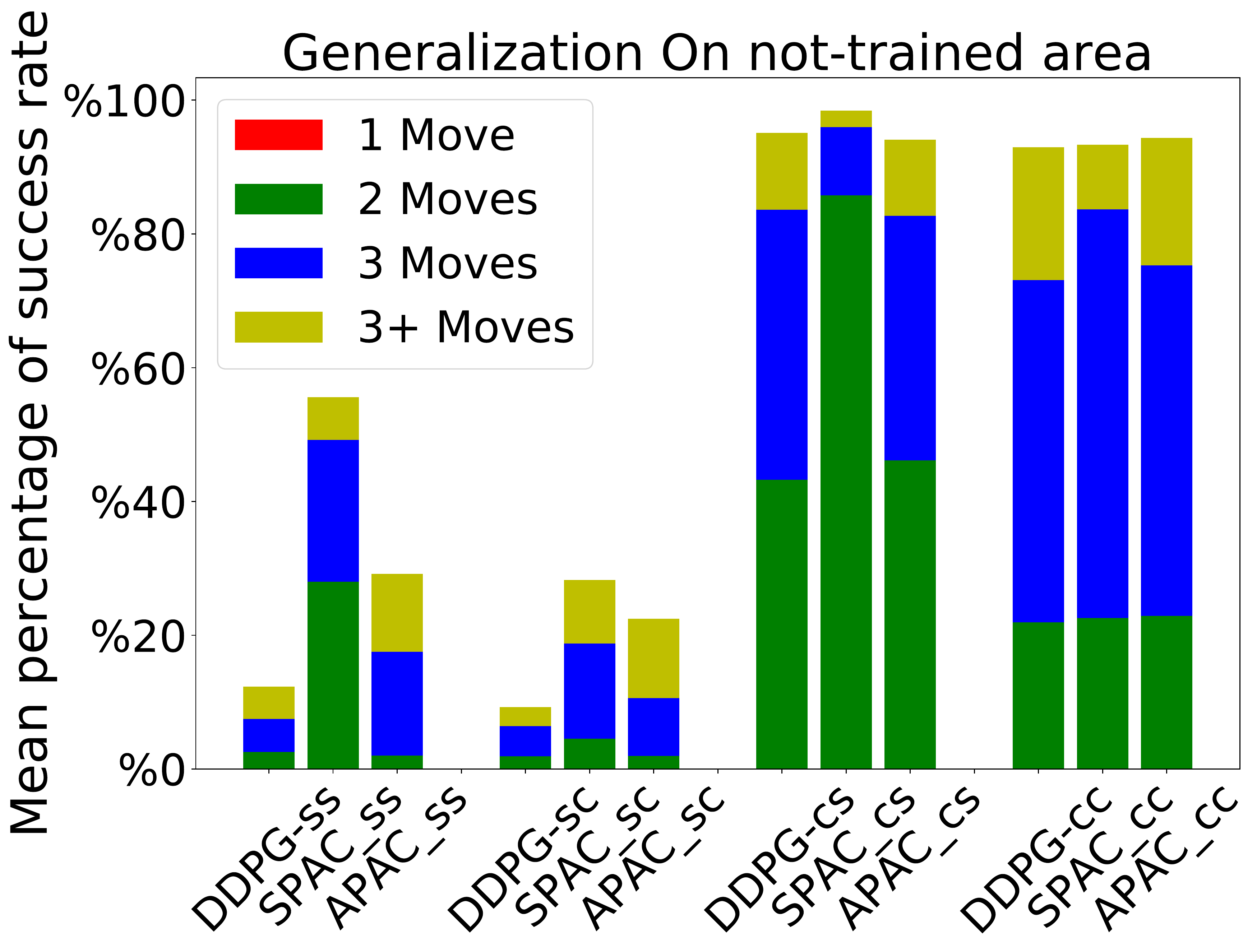}%
	\caption{{\bf Left:} The reaching area for arm under static kinematics conditions. The red area is used for training while the blue region is for testing. Around edges are more coloured since these points can be reached with many more sets of angles. {\bf Right:} Performance of each model is shown under different conditions to reach about 40 targets located in the blue region. All models obtain a good generalization under changing target conditions. }%
	\label{generalization_area}%
\end{figure*}

We also tested the generalization of each model to form of generalization where a whole area of the target zone was not seen during training. This is a form of extrapolation compared to the interpolation trials in the previous generalization experiments. More specifically, we trained each model to reach the target located at a specific region in the environment and we tested each model to reach targets that are located on the unseen area of the environment (see the left most plot in the Figure \ref{generalization_area}). The same distribution of target locations has been used here and only those that are located in the blue area are set as targets for this experiments. Therefore there are about 39 targets under static kinematics conditions and 42 targets under changing kinematics conditions (because of changing kinematics more targets will locate in the testing area). The results show that under static target training, neither model can reach even half of the targets. Their performance is worse under static target and changing kinematics. However, under changing target conditions, all models have obtained a good generalization but they need to take more than one step to reach any target. SPAC has again the best performance among other models under all conditions, while DDPG has the worst performance compared to other two models. These results indicate that learning with a static target hinders generalization as the learner overfits this specific target location.

\begin{figure*}[!t]
\centering
  \includegraphics[width=0.23\textwidth]{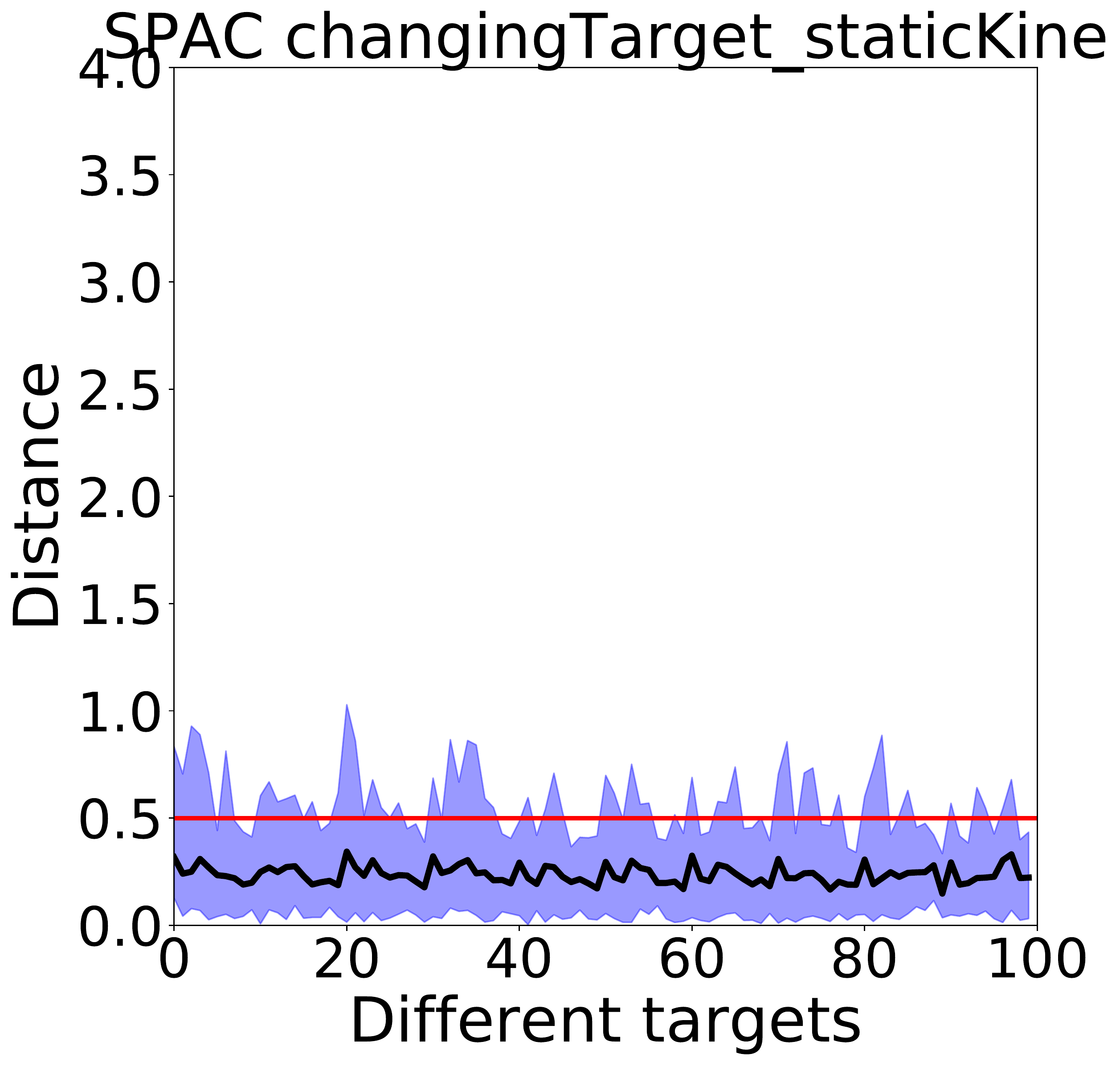}%
  \includegraphics[width=0.23\textwidth]{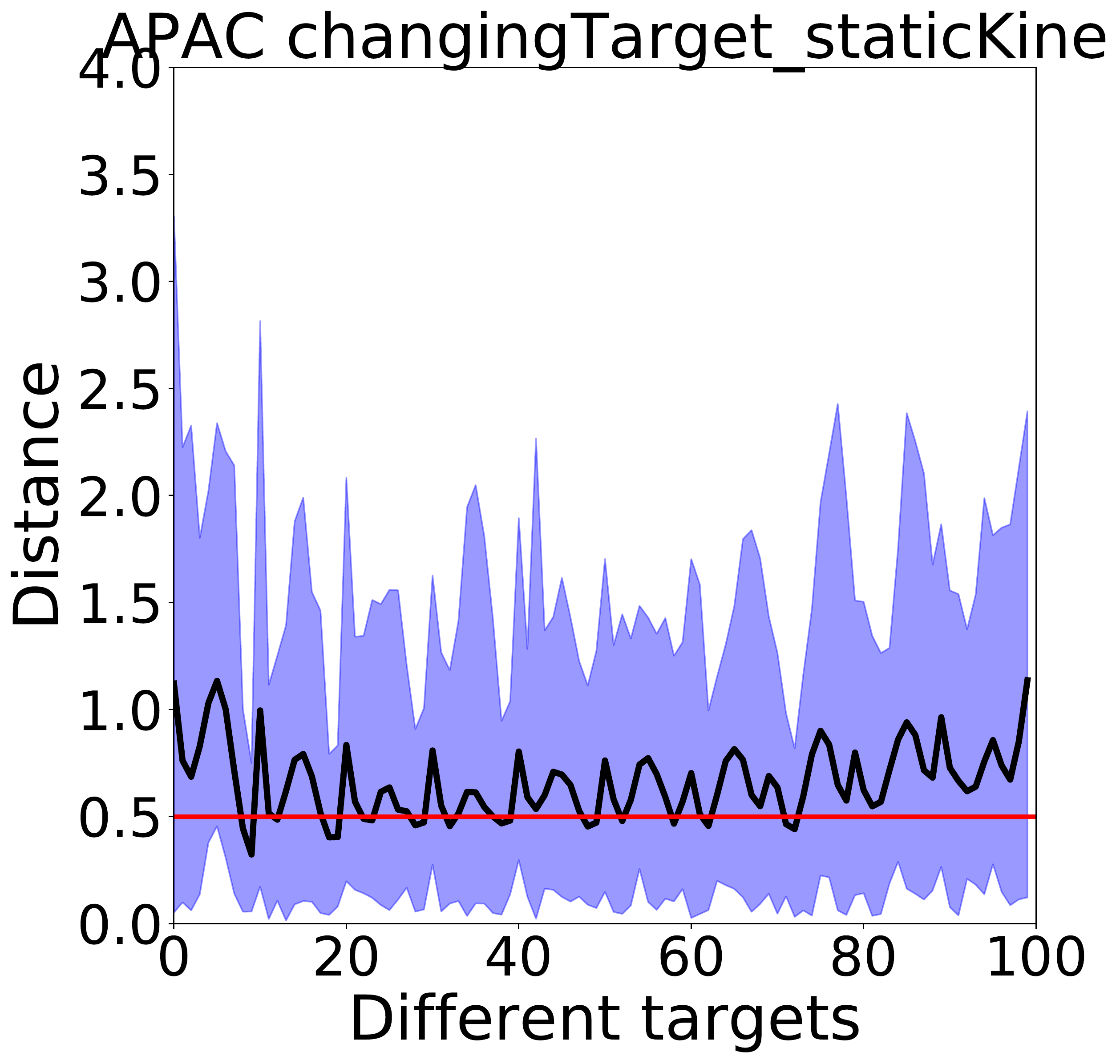}%
  \includegraphics[width=0.24\textwidth]{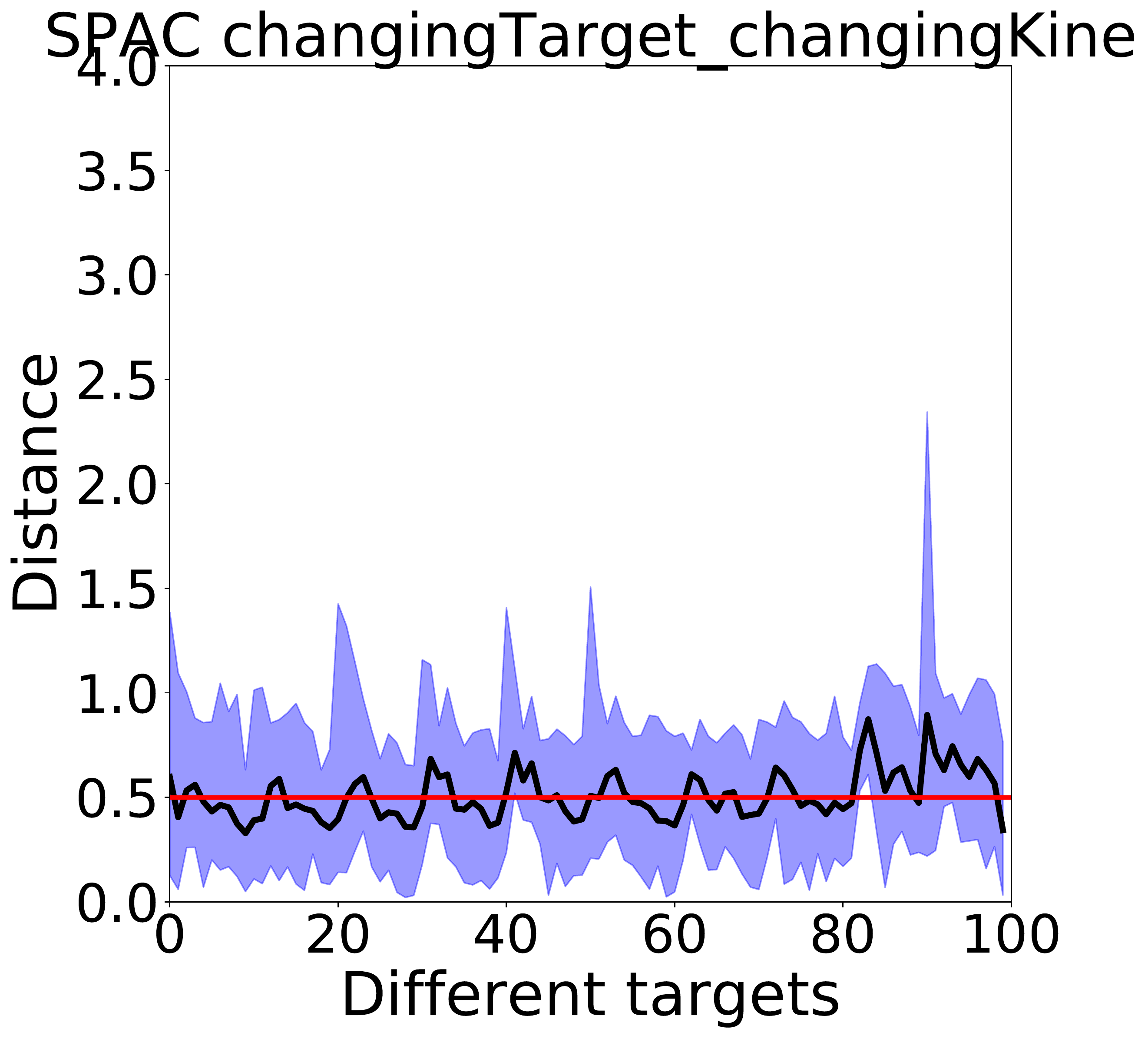}%
  \includegraphics[width=0.24\textwidth]{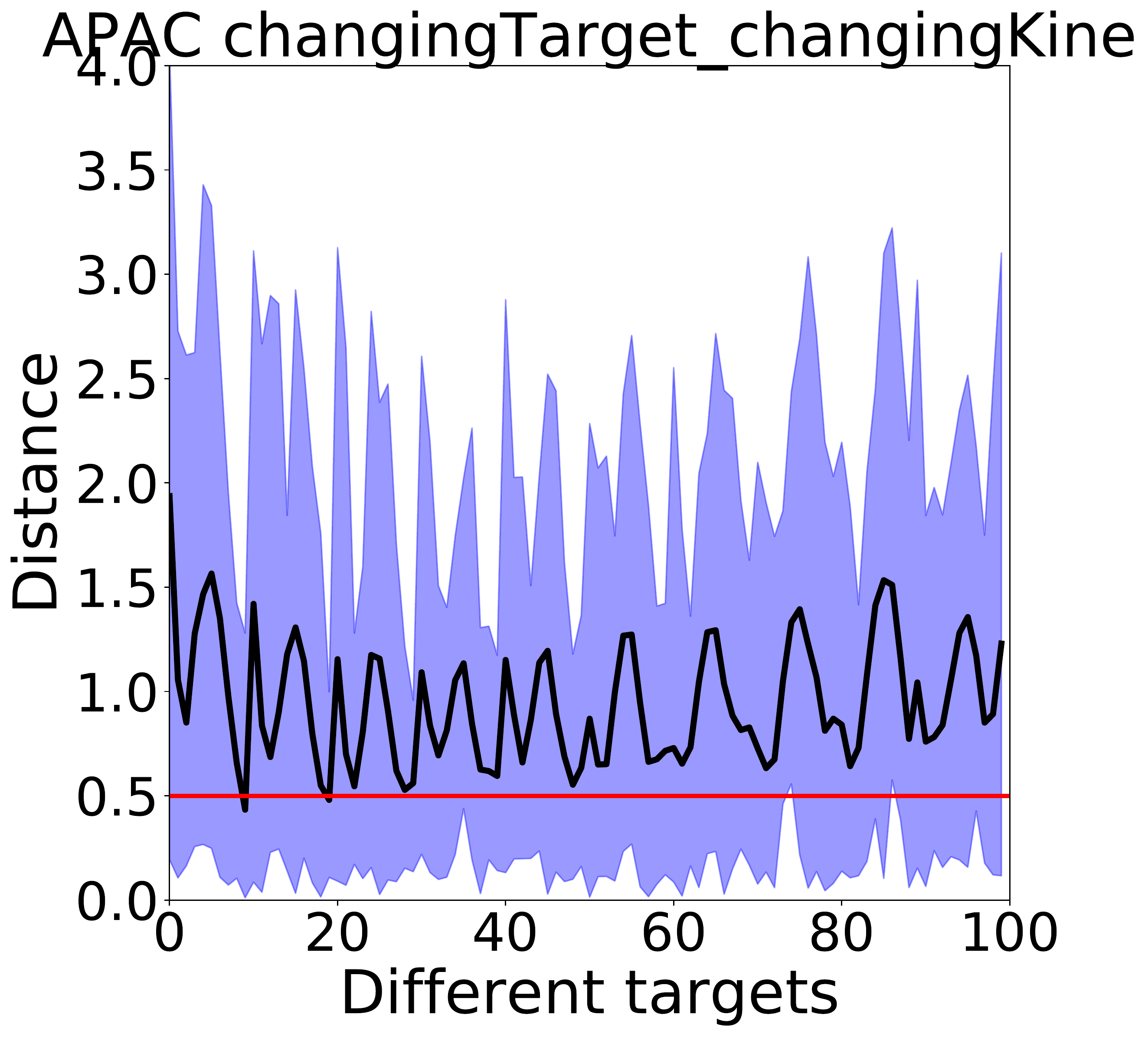}%
  
    \caption[comparison]{The actual distance from the agent to the target location under changing target conditions are shown for SPAC and APAC to reach 100 targets after training over 50 runs. The black line shows the average distance to the target. The area between minimum distance to the maximum distance is coloured in blue. The red line indicates the target zone. In general, SPAC performs better under the dark compared to APAC. }%
   \label{comparison-occluded}%
\end{figure*}

Finally, we want to show results with occluded vision. Since the habitual controller (DDPG) requires sensory input at all times, only the SPAC and APAC models were compared under this condition. 
In these experiments, the arm moves toward the target when the target location is only visible at the first step. When the forward model indicates that the agent has reached the target it stops and the actual distance from the agent (here the arm) is measured. Results of these occluded vision test are summarized in Figure \ref{comparison-occluded} under changing target conditions, since the performance of both models under static target conditions are near perfect.
 
The target zone is marked by the red line in each plot, while the average distance from the agent to the target location is drawn as a black line. The blue area shows the range of the distances that the arm has experienced during the occluded trials when the forward model thinks that it has reached the target. SPAC shows better performance under occluded vision compared to APAC under all conditions. The average actual distance of the end-effector to the target location under changing target/ static kinematic with SPAC is only about a distance of 0.4, which is less than the target zone radius. However, the APAC model under the same conditions stays about a distance of 1 away from the target. Under changing target/changing kinematics, this average actual distance from end-effector to the target is around 0.7 for SPAC and about 1.5 for APAC. It thus seems that any form of habit should be suppressed by a more advanced arbitrator.

\section{Conclusion}

This paper is about the study of a hybrid system with deliberate planning system and habitual control. Habitual control will, of course, be very good after long training in static environments. It was hence important to study the model in changing environments. We investigated the behaviour of our proposed model (APAC) under changing target conditions (to manipulate environmental reward function), changing kinematics of the agent (to manipulate the learned transitional model), and with and without vision. We also tested the model under various conditions to see how good they can generalize their learning paradigm.  The main results are classified as below:

\paragraph{Adaptive to changes:}
Results show significant improvement in performance when planning is available compared to the pure habitual system under changing the environment including changing in the environmental reward function and kinematics of the agent. In comparison, SPAC and APAC are flexible under these changes. This experiment shows that having an internal model is a key to robustness on changing environment. 

\paragraph{Moving from planning to habits:}
By considering that planning is costly, having another control system that is able to provide cheaper solutions can be useful. In this paper, there is no inherent time constraint or computation time difference between the habitual and planning controller. However, if we take this time constraint into account, the APAC is a better model than pure planning (SPAC). Indeed, the overall number of actions taken based on a planned action is less timely in the arbitrated model than the number of planned actions taken by the non-arbitrated model when considering some cost of planning. 

\paragraph{Reaching under occluded vision:}
Since DDPG has no internal model, therefore it can not use any sort of planning to move under occlusion. However, systems like SPAC and APAC build an internal model of the environment that enables them to anticipate and plan a target even when there is no visual information available. Results show that SPAC can perform better in the dark. This is a good example to show that a more sophisticated arbitrator could take different circumstances into account. For occluded vision, such an arbitrator should suppress habitual actions.

Our application example focused on the implementation of the habitual system as an actor-critic for reinforcement learning and a planning system with a forward and inverse model using supervised learning. There have been previous examples of combining both, some form of supervised learning with RL systems and the use of the internal model. In particular, Dyna-Q \cite{sutton1991dyna} and supervised actor-critic \cite{rosenstein2002supervised, barto2004j} are examples of models that bring the capacity of planning into a model-free reinforcement learning space.  
For example, the supervised actor-critic \cite{rosenstein2002supervised, barto2004j} is able to tune the actor manually and very fast when it is needed. This solution is beneficial when dynamics of the system changes dramatically or a new policy is needed to be learned in a very short time. These authors used a gain scheduler that weights the control signal provided by the actor from the reinforcement learner and the supervised actor. In contrast to supervised actor-critic, our proposed model autonomously learns the internal model and arbitrates between the two controllers automatically and not manually. The intention of our model is to study the interaction of habitual and planning systems in form of an arbitrator and ultimately to understanding human behaviour.

Sutton proposed Dyna-Q that is an integrated model for learning and planning\cite{sutton1991dyna}. With respect to our model, Dyna-Q is also a blend of model-free and model-based reinforcement learning algorithms. Dyna-Q can build a transition function and the reward function by hallucinating random samples. Therefore, although it uses a model-free paradigm at the beginning it becomes a model-based solution by learning the model of the world using the hallucination.
In our model, the internal model is used to predict the future state of the agent unlike Dyna-Q that uses the model to train the critic and anticipate the future reward. Moreover, Dyna-Q starts from model-free controller and becomes a model-based controller. Hence, while Dyna-Q has focused on the utilization of internal models to learn a reinforcement controller, our model and study here is concerned with the arbitration of two control systems. 

However, since APAC tend to select actions from the planning controller that it learns very fast, it provides more accurate samples in the experience replay memory. Therefore, similar to the Dyna-Q, the habitual system takes advantage of learning from more valuable samples that lead the habitual controller to a better performance compared to the time that it is trained stand alone (in pure habitual paradigm). 

Not only have human decision-making studies supported the notion that both habitual and planning controls are used during decision-making \citep{daw2011model, ODoherty2015}, there is evidence that arbitration may be a dynamic process involving specific brain regions \cite{lee2014neural}. Our APAC model results suggest that such an arbitration strategy, wherein the planning paradigm is used until the habitual system's predictions become reliable can result in performance that is non-inferior to exclusive planning control in most cases. Thus, our APAC model supports (A) the importance and value of implementing predominantly planning control early in behavioural learning and (B) the diminishing importance of planning control with greater experience in a relatively static environment. 

\section{Acknowledgments}
The authors would like to thank Abraham Nunes for valuable input and acknowledge funding from Natural Science and Engineering Research Council of Canada (NSERC) and NS graduate scholarship.

\bibliographystyle{unsrt}
\bibliography{myBibliography}

\section*{Appendix 1: Algorithms}
To move the arm, a motion function is used that is described in Algorithm \ref{alg:trigonometricFunc}. 
Algorithm APAC is also shown in Algorithm \ref{alg:proposedAlg2}. This algorithm shows how APAC combines the habitual and planning through arbitration. 

\begin{algorithm}[h]
   \caption[Motion Function]{Motion function: function that calculates location of the simulated arm with respect to the applied angles to shoulder $(\alpha)$ and elbow $(\beta)$. $l_1$ and $l_2$ are current length of forearm and arm while $O$ indicates the origin of the plane, where the arm is attached. }
   \label{alg:trigonometricFunc}
\begin{algorithmic}

 \STATE set up an input vector including $(\alpha,\beta,l_1,l_2,O)$
 \STATE $\alpha_1 = \alpha \times \frac{\pi}{180}$
 \STATE $\beta_1 = \beta \times \frac{\pi}{180}$
 \STATE $elbow_x = \cos(\alpha_1) \times l_1$
 \STATE $elbow_y = \sin(\alpha_1) \times l_1$
 \STATE $end_x = elbow_x + \cos(\beta_1+\alpha_1) \times l_2$
 \STATE $end_y = elbow_y + \sin(\beta_1+\alpha_1) \times l_2$
 \STATE return$([end_x+O[0],end_y+O[1]],[elbow_x+O[0],elbow_y+O[1]])$
  \end{algorithmic}
  \end{algorithm}

\begin{algorithm}[]
\begin{spacing}{.8}

   \caption[APAC]{Arbitrator Predictive Actor Critic}
   \label{alg:proposedAlg2}
\begin{algorithmic}
 
 \STATE Randomly initialize critic network $Q(s,a|\theta^Q)$ and actor $\pi(s|\theta^\pi)$ with weights $\theta^Q$ and $\theta^\pi$
   \STATE Initialize target network $Q'$ and $\pi'$ with weights $\theta^{Q'} \leftarrow \theta^Q$, $\theta^{\pi'} \leftarrow \theta^\pi$
   \STATE Randomly initialize forward learner network $fl(s,a|\theta^{f_F})$ and inverse learner network $f_I(s|\theta^{f_I})$ with weights $\theta^{f_L}$ and $\theta^{f_I}$
   \STATE Initialize target network $Q'$ and $\pi'$ with weights $\theta^{Q'} \leftarrow \theta^Q$, $\theta^{\pi'} \leftarrow \theta^\pi$
   \STATE Initialize replay buffer $R$
   \FOR{episode=1,M}
   \STATE Initialize a random process $N$ for action exploration
   \STATE Receive initial observation state $s_1 = [X^{end}_{t},X^{ellbow}_{t},X^{target}]$
   \FOR{t=1,T}
   \IF{episode\textless 100} \STATE {Select action $a_t=f_I(s_t|\theta^{f_I})+N_t$ by the inverse learner} \ELSE 
   \STATE{Compute RPE
   \IF{RPE\textless 1} \STATE {Select action $a_t=\pi([X^{end}_{t},X^{ellbow}_{t},X^{target}]|\theta^{\pi})+N_t$ by the actor} \ELSE 
   \STATE{Select action $a_t=f_I([X^{end}_{t},X^{ellbow}_{t},X^{target}]|\theta^{f_I})+N_t$ by the inverse model} \ENDIF} \ENDIF
   \STATE Feed $a_t$ and current location of the arm to the forward model and observe predicted location of the arm $[X'^{end}_{t+1},X'^{ellbow}_{t+1}]$
   \STATE Execute action $a_t$ and observe reward $r_t$ and observe new location of the arm $[XP^{end}_{t+1},XP^{ellbow}_{t+1}]$
   \STATE Integrate predicted location from forward model and real location and build $s_{t+1}$ along with target location
   \STATE Store transition $(s_t, a_t, r_t, s_{t+1})$ in $R$
   \STATE Sample a random minibatch of $N$ transitions $(s_i, a_i, r_i, s_{i+1})$ from $R$
   \STATE Set $y_i = r_i + \gamma Q'(S_{i+1},\pi'(s_{i+1}|\theta^{\pi'})|\theta^{Q'})$
   \STATE Update forward model by back propagating the error between predicted location and real location of the arm
   \STATE Update inverse model by back propagating the error between predicted action and real taken action by the arm
   \STATE Update critic by minimizing the loss: $L =1/N \sum_i(y_i - Q(s_i,a_i|\theta^Q))^2$
   \STATE Update the actor using policy gradient method:\\ $\nabla{\theta^\pi}J \approx 1/N\sum_i[\nabla_a Q(s,a|\theta^Q)|_{s=s_i,a=\pi(s_i)}\nabla_{\theta^\pi}\pi(s|\theta^\pi)|_{s=s_i}]$
	\STATE Update the inverse model by minimizing difference between action predicted by the inverse model to move  
	 from $X_{t}$ to $X_{t+1}$ with the actual action that transfered the plant from $X_{t}$ to $X_{t+1}$
   \\
   \STATE Update the target networks:\\
    $\theta^{Q'} \leftarrow \tau\theta^Q+(1-\tau)\theta^{Q'}$\\
    $\theta^{\pi'} \leftarrow \tau\theta^\pi+(1-\tau)\theta^{\pi'}$
   
   \ENDFOR
   \ENDFOR 
\end{algorithmic}
\end{spacing}
\end{algorithm}

\section*{Appendix 2: Network architecture, parameters and variables}

It should be again noted that APAC and our proposed predictive actor-critic framework is the extension of DDPG \citep{lillicrap2015continuous}. While preserving almost all parameters and features of the base DDPG model, APAC has added additional networks, which include the forward and inverse models. 

The actor has two fully connected layers with \textit{relu} activation function. These layers are connected to a fully connected output layer with the size of action dimension (here only two neurons). Each of neurons produces a continuous value for either the shoulder or the elbow. Since we need to learn to move the shoulder and elbow in both directions, we selected output activation function to be \textit{tanh} which is multiplied by 180 degrees to obtain -180 to 180 degrees of movements. Note that each joint only moves between 0 to 180 degrees but it can move in two different directions. Actor learning rate is equal to 0.0001. 

The critic has two hidden layers too. The first layer includes 400 fully connected neurons. In the second layer, the output of the actor (300 fully connected neurons) is combined with the output of the first layer (300 fully connected neurons) from the critic to build a fully connected layer with 600 neurons. The output of the critic is only one node since it will learn the value of taken action at each state. All activation functions selected to be \textit{relu}. Learning rate is 0.001. Gamma value to update the TD rule is set to 0.99.

The forward model has three layers of 400, 300, and 4 fully connected neurons. Output layer of this 
network has 4 neurons to predict the future location of the robot arm (end-effector location and elbow location). All activation functions are \textit{sigmoid}. The output of the network is multiplied by the dimension of the environment which is 30 centimetres to produce location of the end-effector and elbow in a Cartesian coordinate. Learning rate is set to 0.1, and the optimization function is "Adamoptimizer" like all other networks.

The inverse model, similar to the actor, has three layers of 400, 300, and 2 fully connected neurons. Output layer of this 
network has 2 neurons to predict proper angles for elbow and shoulder. Activation functions for the first and the second layers are \textit{relu}, and \textit{tanh} is used at the output layer. The output of the network is multiplied by 180 degrees to obtain -180 to 180 degrees of movements similar to the actor.

An experience replay memory of size 1000 samples is used. Each time we train networks using a mini-batch of size 500. 
The smooth update parameter to update target networks is 0.001, and Adamoptimizer is used as an optimization function for actor and critic networks. All the networks use 'L2' regularization with a weight decay by the amount of 0.001.

List of all parameters and variables of all three models (DDPG, SPAC and APAC) are listed here in tables \ref{tab_app1} and \ref{tab_app2}. 

\begin{table}[!ht]
\centering
\caption{{\bf Common parameters for all networks and models}}
\begin{tabular}{l|l}
\hline
\multicolumn{1}{c}{\bf Parameter name}  &\multicolumn{1}{c} {\bf Value}
\\ \hline
Maximum number of trials& 1000\\ 
Maximum number of steps in each trial & 30 \\ 
Discount factor ($\gamma$) & 0.99 \\ 
Soft target network update ($\tau$) & 0.001 \\ 
Random seed for randomizing & None \\ 
Buffer size for experience replay memory & 1000 \\ 
Minibatch size & 500 \\ 
Size of a 2D planar space & $30\times30$ cm \\ 
Origin of the arm fixed on 2D planar space & [15,15] \\ 
Initial length of forearm & 8 cm \\ 
Initial length of arm & 5 cm \\ 
Target zone & area with radius 0.5cm around target\\ \hline

\end{tabular}
\begin{flushleft} 
\end{flushleft}
\label{tab_app1}
\end{table}

\begin{table}[!ht]
\centering
\caption{{\bf Network parameters and variables}}
\begin{tabular}{l|l}
\hline
\multicolumn{1}{c}{\bf Parameters} & \multicolumn{1}{c}{\bf APAC}\\ \hline
Input dimension to actor& 6\\ 
Neurons in the first layer of actor& 400 \\ 
Activation func. in first layer of actor & relu \\ 
Neurons in the second layer of actor& 300 \\ 
Activation func. in second layer of actor & relu \\ 
Neurons in the output layer of actor& 2 \\ 
Scale of output of actor & 180 \\ 
Activation func. in output layer of actor & tanh \\ 
Actor optimization function& Adam optimizer\\ 
Actor learning rate&0.0001 \\ 
\hline
Input dimension to critic& 8\\ 
Neurons in the first layer of critic& 400 \\ 
Activation func. in first layer of critic & relu \\ 
Neurons in the second layer of critic& 600 \\ 
Activation func. in second layer of critic  & relu \\ 
Neurons in the output layer of critic& 1 \\ 
Critic optimization function&Adam optimizer \\ 
Critic learning rate&0.001 \\ 
\hline 
Input dimension to forward model& 6\\ 
Neurons in the first layer of forward model& 400 \\ 
Activation func. in first layer of forward model & sigmoid \\ 
Neurons in the second layer of forward model& 300 \\ 
Activation func. in second layer of forward model & sigmoid \\ 
Neurons in the output layer of forward model& 4 \\ 
Scale of output of forward model & 30 \\ 
Activation func. in output layer of forward model & sigmoid \\ 
Forward model optimization function &Adam optimizer \\ 
Forward model learning rate&0.01 \\ 
\hline
Input dimension to inverse model& 6\\ 
Neurons in the first layer of inverse model& 400 \\ 
Activation func. in first layer of inverse model  & relu \\ 
Neurons in the second layer of inverse model& 300 \\ 
Activation func. in second layer of inverse model  & relu \\ 
Neurons in the output layer of inverse model& 2 \\ 
Scale of output of inverse model & 180 \\ 
Activation func. in output layer of inverse model  & tanh \\ 
Inverse model optimization function&Adam optimizer \\ 
Inverse model learning rate&0.01 \\ 
\hline
\end{tabular}
\begin{flushleft} 
\end{flushleft}
\label{tab_app2}
\end{table}




%
%
%
%
\end{document}